\documentclass[12pt,nofootinbib]{article}
\pdfoutput=1
\usepackage{slashed}
\usepackage{amsmath,amssymb,graphicx,multicol} 
\usepackage{epsf,color}
\usepackage[nosort]{cite}
\usepackage{cancel}
\usepackage{framed}
\usepackage{multirow,array}

\def\hhref#1{\href{http://arxiv.org/abs/#1}{#1}} 

\ifx\pdfoutput\undefined
\usepackage[dvips,bookmarks=false]{hyperref}	
\else
\usepackage{hyperref}	
\fi

\newcommand{\eps}{\epsilon}
\newcommand{\Tr}{\text{Tr}}

\newcommand{\rhoL}{{\rho_L}}
\newcommand{\rhoR}{{\rho_R}}

\begin{document}
\begin{titlepage}
\begin{flushright}
CERN-PH-TH-2015-081
\end{flushright}
\vskip1.3cm

\begin{center}
{\Large \bf 
One-loop effects from spin-1 resonances \\[0.15cm] in Composite Higgs models 
}
\end{center}
\vskip0.7cm

\renewcommand{\thefootnote}{\fnsymbol{footnote}}
\begin{center}
{\large  Roberto Contino$\,^{1,2}$\footnote{\hspace{0.17cm}On leave of absence from Universit\`a
di Roma La Sapienza and INFN, Roma, Italy.} and Matteo Salvarezza$\,^3$
}
\end{center}
\renewcommand{\thefootnote}{\arabic{footnote}}

\vspace{0.1cm}
\begin{center}
{\it 
$^1\,$Institut de Th\'eorie des Ph\'enomenes Physiques, EPFL, Lausanne, Switzerland \\
$^2\,$Theory Division, CERN, Geneva, Switzerland \\
$^3\,$Dipartimento di Fisica, Universit\`a di Roma ``La Sapienza'' and INFN, Roma, Italy
} \\
\vspace*{0.1cm}
\end{center}

\vglue 1.0truecm

\begin{abstract}
\noindent 
We compute the 1-loop correction to the  electroweak  observables from spin-1 resonances in  $SO(5)/SO(4)$ composite Higgs models.
The strong dynamics is modeled with an effective description comprising the Nambu-Goldstone bosons and the lowest-lying spin-1 resonances.
A classification is performed of the relevant operators including custodially-breaking  effects from the gauging of hypercharge.
The 1-loop contribution of the resonances is extracted in a diagrammatic approach by matching to the low-energy theory of Nambu-Goldstone bosons.
We find that the correction is numerically important in a significant fraction of the parameter space and tends to weaken the bounds providing a negative
shift to the $S$ parameter.
\end{abstract}

\end{titlepage}

\section{Introduction}
\label{sec:Intro} 

The electroweak precision measurements performed at LEP, SLD and Tevatron have provided a powerful test of the Standard Model (SM)
and set tight constraints on generic models of new physics. 
They represent a challenge especially for theories where electroweak symmetry breaking (EWSB) originates from new strong dynamics at the TeV scale.
Composite Higgs models~\cite{Kaplan:1983fs,compositeHiggs} are currently the most interesting representative of this class of theories, 
as they can accommodate naturally a light Higgs boson.
The experimental information on  universal corrections to the precision observables at the $Z$ pole can be conveniently summarized in terms of
the three $\eps$ parameters~\cite{Altarelli:1990zd,Altarelli:1991fk}, whose measured value is of order  $\text{a few}\times 10^{-3}$ with an error of $10^{-3}$.
A first important correction to the $\eps_i$ in composite Higgs models arises as a consequence of the modified couplings of the Higgs to the $W$ and $Z$ 
bosons~\cite{Barbieri:2007bh}.
The largest effect comes in particular from the imperfect cancellation of the logarithmic divergence  between 1-loop diagrams with Higgs and EW vector bosons.
The residual divergence, absent in the~SM,  can be interpreted as the running of local effective operators between the scale of new physics $m_\rho$ 
and the electroweak (EW) scale.
This leads to a shift to the~$\eps_i$ which is naively of order $m_W^2/(16\pi^2 f^2) \log (m_\rho/m_Z) \sim 1\times 10^{-4} \, (\xi/0.1) \log (m_\rho/m_Z)$, 
where $f$ is the decay constant of the composite Higgs and $\xi \equiv v^2/f^2$.
Besides the running, a second effect comes from threshold corrections.
Those at the EW scale are model independent; they have been computed in Ref.~\cite{Orgogozo:2012ct} and are small (of order $\text{a few} \times 10^{-5} \, (\xi/0.1)$).
Threshold corrections at the new physics scale $m_\rho$ are instead large, as resonance exchange can give a tree-level contribution to the~$\eps_i$.
In this case one naively expects shifts  
of order $m_W^2/m_\rho^2$, so that a per mille precision on the $\eps_i$ implies a lower limit on $m_\rho$ at the $2-3\,$TeV level.
Given the experimental accuracy, one-loop corrections from the resonances can also give an important contribution. Compared to the IR running
they are subleading by a factor $\log(m_\rho/m_Z)$, although  this latter is  numerically not very large in natural scenarios 
(e.g. $\log(m_\rho/m_Z) \simeq 3.6$ for $m_\rho =3\,$TeV) and can be compensated by a multiplicity factor from the loop of resonances or simply 
by a numerical accidental enhancement. For example, one-loop corrections from fermionic resonances to $\eps_3$ are enhanced by color and generation
multiplicity factors~\cite{Grojean:2013qca,Azatov:2013ura}, while those to $\eps_1$ represent the leading effect from new physics if the strong dynamics is 
custodially symmetric~\cite{dTfromfermions,Giudice:2007fh,Barbieri:2007bh,Grojean:2013qca}.

Aim of this work is to compute the one-loop threshold corrections due to spin-1 resonances in composite Higgs models. 
These effects were studied in detail in the framework of strongly-interacting Higgless models (with an $SO(4)/SO(3)$ coset), 
for which computations exist both in the diagrammatic approach~\cite{Matsuzaki:2006wn,Barbieri:2008cc,Cata:2010bv,Foadi:2012ga} 
and through the use of dispersion relations~\cite{Orgogozo:2011kq,Pich:2012jv}.
Previous analyses of composite Higgs models, on the other hand,  included the contribution of spin-1 resonances only at the tree level, see for example 
Ref.~\cite{Orgogozo:2012ct} 
for a generalization of the Peskin-Takeuchi dispertion relation for the $S$ parameter~\cite{Peskin:1991sw} to $SO(5)/SO(4)$.
In this paper we perform a  calculation of these one-loop threshold effects in $SO(5)/SO(4)$ composite Higgs theories by modeling the strong dynamics with 
a simple effective description including the Nambu-Goldstone (NG) bosons and the lowest-lying spin-1 resonances.
These latter are assumed to be lighter and more weakly interacting than the other composite states at the cutoff.  Although this working assumption might not
be realized by the underlying strong dynamics, we expect our calculation to give a quantitative approximate description of the contributions from spin-1 resonances
arising in full models.
Our results represent a required step towards a complete one-loop analysis of precision observables in composite Higgs models
including both fermionic and bosonic resonances.

This paper is organized as follows. Section~\ref{sec:effeLagrangian} discusses the effective Lagrangian for the NG bosons and the spin-1 resonances, 
highlighting the role of symmetries. The computation of the one-loop correction to the $\epsilon$ parameters from spin-1 resonances is illustrated
in Section~\ref{sec:calculation}. The heavy states are integrated out at a scale $\mu \sim m_\rho$ matching 
to the low-energy theory with only NG bosons.
Our results are used to perform a fit to the electroweak observables in Section~\ref{sec:fit}, where limits on the scale $m_\rho$ and the degree of Higgs compositeness  $\xi$
are derived. We draw our conclusions in Section~\ref{sec:conclusions}. Finally, we collect in the Appendices some useful additional results: 
Section~\ref{app:two-site} discusses 
the two-site limit of the spin-1 Lagrangian; Sections~\ref{app:formulas},~\ref{sec:alpha2oneloop}~and~\ref{sec:singlerho} report formulas related 
to our calculation;  a discussion of the one-loop renormalization of the spin-1 
Lagrangian is given in Section~\ref{sec:spin1renorm}; while Section~\ref{sec:altmatchingcT} provides an alternative derivation of the matching for the $T$ parameter.

\section{Effective Lagrangian and its symmetries}
\label{sec:effeLagrangian} 

We construct the low-energy  effective Lagrangian describing the NG bosons and massive \mbox{spin-1} resonances by using the formalism 
of Callan, Coleman, Wess and Zumino (CCWZ)~\cite{Coleman:1969sm} 
for $SO(5)/SO(4)$. We  follow closely the notation of Refs.~\cite{Contino:2011np,Azatov:2013ura},  to which we refer the reader for more details. 
Nambu-Goldstone bosons are parametrized in terms of the field $U(\pi) = \exp(i \sqrt{2} \pi(x)/f)$, where $\pi(x) = \pi^{\hat a}(x) T^{\hat a}$ and $f$ is
the associated decay constant.~\footnote{We denote with $T^{a} = \{ T^{a_L} , T^{a_R} \}$ the generators of $SO(4) \sim SU(2)_L \times SU(2)_R$ 
and with $T^{\hat a}$ those of $SO(5)/SO(4)$, normalized such that $\Tr [T^A T^B] = \delta^{AB}$.} 
Under global rotations $g \in SO(5)$, the NG fields transform as
\begin{equation}
U(\pi) \to U(g(\pi)) = g \, U(\pi) h^\dagger(g,\pi)\, ,
\end{equation}
where $h(g,\pi(x))$ is an  element of $SO(4)$ which depends on $g$ and $\pi(x)$.
The CCWZ construction makes use of the covariant functions $d_\mu(\pi) = d_\mu(\pi)^{\hat a} T^{\hat a}$ and 
$E_\mu^L(\pi) = E^{a_L}(\pi) T^{a_L}$, $E_\mu^R(\pi) = E^{a_R}(\pi) T^{a_R}$, which are defined by
\begin{equation} \label{eq:defdE}
-i U^\dagger(\pi) D_\mu U(\pi) = d_\mu(\pi) + E_\mu^L(\pi) + E_\mu^R(\pi)
\end{equation}
and transform as 
\begin{equation}
\begin{split}
d_\mu(\pi) & \to h(g,\pi) d_\mu(\pi) h^\dagger(g,\pi) \\
E_\mu(\pi) & \to h(g,\pi) E_\mu(\pi) h^\dagger(g,\pi) - i  h(g,\pi) \partial_\mu  h^\dagger(g,\pi) \, .
\end{split}
\end{equation}
In particular, $E_\mu= E^L_\mu + E^R_\mu$ transforms as a gauge field of $SO(4)$ and can be used to define a covariant derivative 
$\nabla_\mu = \partial_\mu + i E_\mu$  as well as a field strength $E_{\mu\nu} = \partial_\mu E_\nu -  \partial_\nu E_\mu + i [E_\mu , E_\nu]$.
The SM electroweak vector bosons weakly gauge a subgroup $SU(2)_L \times U(1)_Y \subset  SO(4)'$ contained in $SO(5)$, 
where the $SO(4)'$ is misaligned by an angle $\theta$ with respect to the unbroken $SO(4)$. Hypercharge is identified with $Y = T^{3_R}_0$, where
$T^{a_L}_0$, $T^{a_R}_0$ are the generators of $SO(4)'$~\footnote{They are related to the  generators $\{ T^a , T^{\hat a}\}$ through a rotation by an angle $\theta$:
$T^A_0 = r^{-1}(\theta) T^A r(\theta)$, see Ref.~\cite{Contino:2011np}.}. The derivative appearing in Eq.~(\ref{eq:defdE}) is thus covariant with respect to 
local transformations of $SU(2)_L \times U(1)_Y$: $D_\mu = \partial_\mu + i W_\mu^{a_L} T_0^{a_L} +i B_\mu Y$.
Although the EW gauging introduces an explicit breaking of the global $SO(5)$ symmetry,
the low-energy Lagrangian can still be expressed in an $SO(5)$-invariant fashion by introducing suitable spurions that encode the breaking. 
We will be mainly interested
in custodially-breaking radiative effects induced by loops of the hypercharge field, while $W_\mu$ will be treated as an external source. In this limit the explicit
breaking of $SO(5)$ can be parametrized in terms of a single spurion
\begin{align}
& \chi(\pi)  = U^\dagger(\pi) g' T_0^{3_R} U(\pi)\, ,
\intertext{whose formal transformation rule is}
& \chi  \to h(g,\pi)\, \chi\, h^\dagger(g,\pi)\, .
\end{align}

The part of the Lagrangian which describes the interactions among NG bosons can be organized in a derivative
expansion controlled by  $\partial/\Lambda$:
\begin{equation} \label{eq:effLpion}
{\cal L}(\pi) = {\cal L}^{(2)}(\pi) + {\cal L}^{(4)}(\pi) + {\cal L}^{(6)}(\pi) + \dots 
\end{equation}
where $\Lambda\lesssim 4\pi f$  is the cutoff of the effective theory and ${\cal L}^{(n)}$ indicates terms with $n$ derivatives.
Omitting for simplicity $CP$-violating operators, one has:~\footnote{Additional $O(p^2)$ operators with two powers
of the spurion  are not linearly independent.
Specifically, by using the identity $\nabla_\mu \chi = -i [d_\mu,\chi]$ it is easy to  show that:
\begin{equation}
\begin{split}
\Tr\!\left[ \nabla_\mu \chi \nabla^\mu \chi \right] & = 2\, \Tr\!\left[ d_\mu d^\mu \chi^2 \right]  - \left( \Tr\!\left[d_\mu \chi \right] \right)^2 \\
\Tr\!\left[ d_\mu \chi d^\mu \chi \right] & = \frac{1}{2} \left( \Tr\!\left[d_\mu \chi \right] \right)^2 \, .
\end{split}
\end{equation}
}
\begin{align}
\label{eq:Lpi}
{\cal L}^{(2)}(\pi)& = \frac{f^2}{4} \Tr\!\left[d_\mu d^\mu \right] +   c_{T} \, f^2\! \left( \Tr\!\left[d_\mu \chi \right] \right)^2 
+   c_{\chi} \, f^2 \Tr\!\left[d_\mu d^\mu \chi^2 \right] \\[0.3cm]
{\cal L}^{(4)}(\pi)& = \sum_i c_i O_i + \dots 
\end{align}
where
\begin{equation}
\label{eq:CCWZbasis}
\begin{split}
O_1 & = \Tr\!\left[ d_\mu d^\mu \right]^2 \\[0.1cm]
O_2 & = \Tr\!\left[ d_\mu d_\nu \right] \Tr\!\left[ d^\mu d^\nu \right] \\[0.1cm]
\end{split}
\qquad\quad
\begin{split}
O_3^\pm & = \Tr\!\left[ (E^L_{\mu\nu})^2 \pm (E^R_{\mu\nu})^2 \right] \\[0.1cm]
O_4^\pm & = \Tr\!\left[ \left(E^L_{\mu\nu} \pm E^R_{\mu\nu}\right) i [d^\mu , d^\nu] \right] 
\end{split}
\end{equation}
and the dots stand for higher-derivative terms and $O(p^4)$ operators involving $\chi$.
We adopted the basis of four-derivative $SO(5)$-invariant operators of Ref.~\cite{Azatov:2013ura} (see also Ref.~\cite{Contino:2011np}) but dropped the operator 
$O_5$ there appearing because it identically vanishes~\cite{Alonso:2014wta}.
Among the terms with 6 derivatives we only list two operators that are relevant for our analysis:
\begin{equation}
{\cal L}^{(6)}(\pi) = c_{2W} \left( \nabla^\mu E_{\mu\nu}^L \right)^2 + c_{2B} \left( \nabla^\mu E_{\mu\nu}^R \right)^2  + \dots
\end{equation}

The operators $O_3^-$, $O_4^-$ are odd under the action of the parity $P_{LR}$ exchanging the $SU(2)_L$ and $SU(2)_R$ groups inside the unbroken 
$SO(4)$~\cite{Contino:2011np}; all the other operators in Eqs.~(\ref{eq:Lpi}),(\ref{eq:CCWZbasis}) are $P_{LR}$ even. In particular, under $P_{LR}$ 
the spurion $\chi$ transforms as 
\begin{equation}
\chi \to P_{LR}\, U^\dagger(\pi) g' T^{3_L}_0 U(\pi) P_{LR}\equiv P_{LR} \, \tilde \chi \, P_{LR} \, .
\end{equation}
Considering that $\Tr[ d_\mu \tilde\chi ] = - \Tr[ d_\mu \chi ]$ and $\chi^2 = \tilde\chi^2$, 
it easily follows that the operators $O_{T} = f^2\! \left( \Tr\!\left[d_\mu \chi \right] \right)^2$ and $O_{\chi} = f^2 \Tr\!\left[d_\mu d^\mu \chi^2 \right]$  are
even under $P_{LR}$. While $O_{\chi}$ is also custodially symmetric,~\footnote{The operator $O_\chi$ breaks explicitly $SO(5)$ down to the gauged $SO(4)'$.
This can be easily seen by rewriting $\Tr\!\left[d_\mu d^\mu \chi^2 \right] = \Tr\!\left[d_\mu d^\mu \right] - (U d_\mu d^\mu U^\dagger)_{55}$, where
the gauged $SO(4)'$ acts on the first four components of $SO(5)$. In the unitary gauge one has 
$O_\chi = (f^2/16) [(W^1_\mu)^2 + (W^2_\mu)^2 + (B_\mu - W^3_\mu)^2] \sin^2(\theta+h/f) (1+\sin^2(\theta+h/f))$, which is custodially symmetric.} 
the operator $O_{T}$ is the only one which breaks explicitly the custodial symmetry and thus contributes to the $T$ parameter.
The $S$ parameter instead gets a contribution from $O_3^+$~\cite{Contino:2011np,Azatov:2013ura}~\footnote{In the unitary gauge (with gauge kinetic
terms normalized as $-W_{\mu\nu}^{a} W^{\mu\nu\, a}/4g^2 $, $- B_{\mu\nu} B^{\mu\nu}/4g'^2$)
\begin{equation}
\begin{split}
O_{T}\big|_{u. gauge} & = \frac{g'^2 f^2}{4} \sin^4\!\left(\theta + \frac{h}{f} \right) \left( W_\mu^3 - B_\mu\right)^2 \\[0.1cm]
O_{3}^+\big|_{u. gauge}  & = \frac{1}{2} \sin^2\!\left(\theta + \frac{h}{f} \right) \left( (W_{\mu\nu}^a)^2 + (B_{\mu\nu})^2 - 2 W_{\mu\nu}^3 B^{\mu\nu} \right) + \dots
\end{split}
\end{equation}
where the dots indicate terms with more than two gauge fields. By expanding in powers of the fields, at the level of dimension-6 operators, one has
\begin{equation}
\begin{split}
O_T & = \frac{g'^2}{f^2} | H^\dagger  {\overleftrightarrow { D_\nu}} H |^2 + \dots \\
O_3^+ & = - \frac{i}{2 f^2} D^\nu W_{\mu\nu}^i (H^\dagger  \sigma^i  {\overleftrightarrow { D^\mu}} H) 
                   - \frac{i}{2 f^2} \partial^\nu B_{\mu\nu} (H^\dagger  {\overleftrightarrow { D^\mu}} H) + \dots
\end{split}
\end{equation}
}.

Spin-1 resonances will be described by vector fields 
$\rho^L_\mu = \rho_\mu^{a_L} T^{a_L}$ and $\rho^R_\mu = \rho_\mu^{a_R} T^{a_R}$ living
in the adjoint of $SO(4)\sim SU(2)_L \times SU(2)_R$ and transforming non-homogeneously under $SO(5)$ global rotations:
\begin{equation}
\rho_\mu \to h(g,\pi) \rho_\mu h^\dagger(g,\pi) - i  h(g,\pi) \partial_\mu h^\dagger(g,\pi)\, .
\end{equation}
We will assume that the Lagrangian that describes their interactions can also be organized in a derivative
expansion controlled by  $\partial/\Lambda$, so that physical quantities at $E\ll \Lambda$ are saturated by the lowest terms~\cite{Contino:2011np}.
In order to estimate the coefficients of the operators appearing in the effective Lagrangian, we adopt the criterion of Partial UV Completion (PUVC)~\cite{Contino:2011np}.
This premises that the coupling strengths of the resonances to the NG bosons and to themselves do not exceed, and preferably saturate, the $\sigma$-model 
coupling $g_* = \Lambda/f$ at the cutoff scale. Under this assumption, neglecting for simplicity $CP$-odd operators, the leading terms in 
the Lagrangian are
\begin{equation} \label{eq:Lrho}
\begin{split}
{\cal L}(\rho) = & \sum_{r=L,R} \bigg\{  
  -\frac{1}{4g_{\rho_r}^2} \,\Tr\!\left(\rho^r_{\mu\nu} \rho^{r\,\mu\nu} \right) + \frac{m_{\rho_r}^2}{2 g_{\rho_r}^2} \,\Tr\!\left( \rho^r_\mu- E_\mu^{r} \right)^2 \\
 & \phantom{\sum_{r=L,R} \bigg\{  \,}
    + \beta_{1r} \,\Tr\!\left[ (\rho^r_\mu- E_\mu^{r} ) \chi \right]\!\Tr\!\left( d^\mu \chi \right) 
    + \beta_{2r} \left(\Tr\!\left[ (\rho^r_\mu- E_\mu^{r} ) \chi \right]\right)^2 \\
 & \phantom{\sum_{r=L,R} \bigg\{  \,}
     + \alpha_{1r} \,\Tr\!\left(\rho^r_{\mu\nu}\, i[d^\mu,d^\nu]\right)  + \alpha_{2r} \,\Tr\!\left(\rho^{r\,\mu\nu} E_{\mu\nu}^r \right)  \bigg\}  \\[0.05cm] 
 & + \beta_{LR} \, \Tr\!\left[ (\rho^L_\mu- E_\mu^{L} ) \chi \right] \Tr\!\left[ (\rho^R_\mu- E_\mu^{R} ) \chi \right] \, .
\end{split}
\end{equation}
Among the operators involving $\chi$, we have kept only those relevant for the present analysis.

\subsection{Hidden local symmetry description}
\label{sec:HLS}

The above construction relies on describing the resonances in terms of massive vector fields, which propagate three polarizations. 
At energies $m_\rho \ll E < \Lambda$, however, the longitudinal and transverse polarizations behave differently  (their interactions scale differently
with the energy), and it is convenient to describe them in terms of distinct fields. Indeed, it is always possible to parametrize the longitudinal 
polarizations of massive spin-1 fields in terms of a set of eaten NG bosons~\footnote{See for example Ref.~\cite{Burgess:1992gz}.}. 
In the case of the Lagrangian~(\ref{eq:Lrho}) 
the corresponding coset is $SO(5) \times SO(4)_H/SO(4)_{d}$, which leads to 10 NG bosons transforming under the unbroken diagonal $SO(4)_d$ as $\pi = (2,2)$,
$\eta^L = (3,1)$ and $\eta^R = (1,3)$~\cite{Contino:2011np}. Their  $\sigma$-model Lagrangian can be obtained 
by taking the limit  $g_\rho \to 0$ with  $m_\rho/g_\rho$  fixed;
transverse polarizations are then reintroduced by gauging the $SO(4)_H$ subgroup with vector fields $\rho_\mu$.
It is  convenient to parametrize the NG bosons in terms of  $U(\pi,\eta) = e^{i \sqrt{2} \pi/f} e^{i\eta^L/f_{\rhoL}} e^{i\eta^R/f_{\rhoR}}$~\cite{Contino:2011np},
where $\eta^L(x) = \eta^{a_L}(x) X^{a_L}$, $\eta^R(x) = \eta^{a_R}(x) X^{a_R}$ and, we recall, $\pi(x) = \pi^{\hat a}(x) T^{\hat a}$~\footnote{We denote the 
$SO(5) \times SO(4)_H/SO(4)_{d}$ (broken) generators by $T^{\hat a},  X^a = (T^a - T^a_H)/\sqrt{2}$, where $T_H^a$ are those of $SO(4)_H$, 
and the $SO(4)_d$ (unbroken) generators by  $Y^a = (T^a + T^a_H)/\sqrt{2}$. We will consider their matrix representation on a $9\times 9$ space, 
so that $T^a$, $T^{\hat a}$ and $T^a_H$ act respectively on $5\times 5$ and $4\times 4$ subspaces. All the traces in this section and in the next one
(Sections~\ref{sec:HLS} and~\ref{sec:twosite}) will be $9\times 9$ ones except where explicitly indicated.
}.
It is thus straightforward to derive the CCWZ decomposition
\begin{align}
-i U^\dagger D_\mu U & = d_\mu(\pi,\eta) + \tilde d_\mu^L(\pi,\eta) + \tilde d_\mu^R(\pi,\eta) + E_\mu^L(\pi,\eta) + E_\mu^R(\pi,\eta)  \\[0.4cm]
\begin{split}
 d_\mu(\pi,\eta) & =  e^{-i\eta^R/f_{\rhoR}}e^{-i\eta^L/f_{\rhoL}}\, d_\mu(\pi)\, e^{i\eta^L/f_{\rhoL}} e^{i\eta^R/f_{\rhoR}} \\
\tilde d_\mu^r(\pi,\eta) + E_\mu^r(\pi,\eta) & = e^{-i\eta^r/f_{\rho_r}} \left( -i \partial_\mu + E^r_\mu(\pi) + \rho_\mu^r \right) e^{i\eta^r/f_{\rho_r}}  \qquad (r = L,R)\, , 
\end{split}
\end{align}
where $d_\mu(\pi,\eta)$, $\tilde d_\mu(\pi,\eta)$ and $E_\mu(\pi,\eta)$ are obtained by projecting respectively along  the generators $T^{\hat a}$, $X^a$ and $Y^a$.
Here $d_\mu(\pi)$ and $E_\mu(\pi)$ denote the uplift of the corresponding $SO(5)/SO(4)$ functions to the $9\times 9$ space (they have non-vanishing components
in the $5\times 5$ subspace). 
Notice that $d_\mu(\pi,\eta)$ is just an ($\eta$-dependent) $SO(4)_d$ rotation of $d_\mu(\pi)$.
Since $SO(5) \times SO(4)_H/SO(4)_{d}$ is not a symmetric space, hence no grading of the algebra exists, the $d$ and $E$ symbols will contain terms with 
both odd and even numbers of NG bosons in their expansion. In particular,
\begin{equation}
\begin{split}
(\tilde d^L_\mu)^{a_L} = & \, \frac{1}{f_\rhoL} \partial_\mu\eta^{a_L} + \frac{1}{\sqrt{2}} \left( E^L_\mu(\pi) - \rho^L_\mu \right)^{a_L} - \frac{1}{2f_\rhoL} \epsilon^{a_L b_L c_L}
 \left( E^L_\mu(\pi) + \rho^L_\mu \right)^{b_L} \eta^{c_L}  \\ 
 & + \frac{1}{4\sqrt{2}f_\rhoL} \left[ \eta^{a_L}  \left( E^L_\mu(\pi) - \rho^L_\mu \right)^{b_L} \eta^{b_L} - 
      \left( E^L_\mu(\pi) - \rho^L_\mu \right)^{a_L}  \eta^{b_L}  \eta^{b_L} \right] + \dots
\end{split}
\end{equation}
and similarly for $\tilde d^R_\mu$. 
In the unitary gauge $\eta^a=0$ one has $(\tilde d^r_\mu)^{a} = ( E^r_\mu(\pi) - \rho^r_\mu)^{a}/\sqrt{2}$ ($r=L,R$).
It is thus easy to see that the kinetic terms of the NG bosons $\eta$ are mapped into the $\rho$ mass terms of Eq.~(\ref{eq:Lrho}),
\begin{equation}
\label{eq:Ukin}
\frac{f_{\rho_r}^2}{2} \,\Tr\big( \tilde d_\mu^r(\pi,\eta) \tilde d^{r\, \mu}(\pi,\eta) \big)\ \longrightarrow \ 
\frac{f_{\rho_r}^2}{4}\,\Tr\!\left[\left( \rho^r_\mu- E_\mu^{r}(\pi) \right)^2\right]_{5\times 5}\, , 
\end{equation}
(where $[ \ \  ]_{5\times 5}$ denotes a $5\times 5$ trace) with the identification
\begin{equation} \label{eq:arho}
a_{\rho_r} \equiv \frac{m_{\rho_r}}{g_{\rho_r} f} = \frac{1}{\sqrt{2}} \frac{f_{\rho_r}}{f} \qquad (r=L,R)\, . 
\end{equation}
At the level of terms quadratic in the $d$ symbols, other three operators with two powers of $\chi$ map into those with coefficients $\beta_i$ in Eq.~(\ref{eq:Lrho}),
once evaluated in the unitary gauge:
\begin{equation} \label{eq:UcountT}
\begin{split}
\Tr\big( \tilde d_\mu^r(\pi,\eta) \chi(\pi,\eta) \big) \Tr\big(  d^\mu(\pi,\eta)  \chi(\pi,\eta) \big)  
   & \longrightarrow \  - \frac{1}{2} \Tr\!\left[ \bar\rho^r_\mu \, \chi(\pi) \right]_{5\times 5}\!\Tr\!\left[ d^\mu(\pi) \chi(\pi) \right]_{5\times 5}
\\[0.4cm]
\big(\Tr\big[ \tilde d_\mu^r(\pi,\eta) \chi(\pi,\eta) \big]\big)^2  & \longrightarrow  \ \frac{1}{4} \left(\Tr\!\left[ \bar\rho^r_\mu\, \chi(\pi) \right]_{5\times 5}\right)^2
\\[0.4cm]
\Tr\big[ \tilde d_\mu^L(\pi,\eta) \chi(\pi,\eta) \big] \Tr\big[  \tilde d_\mu^R(\pi,\eta)  \chi(\pi,\eta) \big] \
   & \longrightarrow \   \frac{1}{4} \Tr\!\left[ \bar\rho^L_\mu \, \chi(\pi) \right]_{5\times 5} \Tr\!\left[ \bar\rho^R_\mu \, \chi(\pi) \right]_{5\times 5} \, .
\end{split}
\end{equation}
Here we defined $\bar\rho^r_\mu \equiv \rho^r_\mu - E^r_\mu(\pi)$ and $\chi(\pi,\eta) \equiv U^\dagger(\pi,\eta) T_0^{3_R} U(\pi,\eta)$.

\subsection{Two-site model limit}
\label{sec:twosite}

While in general $\pi$, $\eta^L$, $\eta^R$ form three irreducible representations of the unbroken group, in the gauge-less limit $g_\rho = g = g' = 0$  and
for the special choice $f_\rhoL = f_\rhoR = f$  the $O(p^2)$ Lagrangian
\begin{equation} \label{eq:kinterms}
\frac{f^2}{4} \,\Tr\big( d_\mu(\pi) d^{\mu}(\pi) \big) 
 + \frac{f_{\rho_L}^2}{2} \,\Tr\big( \tilde d_\mu^L(\pi,\eta) \tilde d^{L\, \mu}(\pi,\eta) \big)
 + \frac{f_{\rho_R}^2}{2} \,\Tr\big( \tilde d_\mu^R(\pi,\eta) \tilde d^{R\, \mu}(\pi,\eta) \big)
\end{equation}
is invariant under a larger $SO(5) \times SO(5)_H\to SO(5)_d$ global symmetry, under which the NG bosons transform as a single representation: a 10 of $SO(5)_d$. 
In this limit Eq.~(\ref{eq:kinterms}) describes an $SO(5) \times SO(5)_H$ two-site model, where the EW vector bosons and the 
$\rho$ gauge respectively the left and right site~\cite{Panico:2011pw}. 
By virtue of 
Eqs.~(\ref{eq:Ukin}) and (\ref{eq:arho}), the same two-site description is obtained from a Lagrangian containing the kinetic and mass terms for $\pi$ and~$\rho$ 
(first term of Eq.~(\ref{eq:Lpi}) and first two terms of Eq.~(\ref{eq:Lrho})) for $a_\rhoL = a_\rhoR = 1/\sqrt{2}$.
Another, more convenient, parametrization of the Nambu-Goldstone bosons is also possible in this case
in terms of a $5\times 5$ link field, $\bar U(\pi,\eta)$, transforming as a $(5,\bar 5)$ of $SO(5) \times SO(5)_H$, see Appendix~\ref{app:two-site}.
As discussed in detail in Ref.~\cite{Panico:2011pw}, the interest of the two-site model lies in the fact that the Higgs boson is doubly protected,
and EWSB effects stem from a collective breaking of the global symmetry. There are indeed two sources of explicit breaking of 
 $SO(5) \times SO(5)_H$:
the EW gauging of an $SU(2)_L \times U(1)_Y$ subgroup of $SO(5)$ on the left site, and the gauging of $SO(4)_H$ by the $\rho$ on the
right site. If either of these two gaugings is switched off, there is an unbroken $SO(5)$ symmetry which allows one to
align the vacuum to $\theta =0$ without loss of generality. 
This means that for $g_\rho \to 0$, with non-zero EW couplings, all EWSB effects must vanish in the two-site model.
Indeed, the Higgs is a NG boson under both $SO(5)$'s, and both symmetries must be  explicitly broken (hence the collective breaking) 
in order to generate any EWSB effect. 

The authors of Ref.~\cite{Panico:2011pw} also put forward a simple power counting argument showing that collective
breaking lowers the superficial degree of divergence of EWSB quantities. 
This is easy to see by working in a renormalizable
gauge and noticing that the NG bosons $\eta$ interact with  strength $E/f_\rho$, while the gauge field $\rho_\mu$ has coupling~$g_\rho$.  In any 1PI diagram,  
replacing  an internal $\eta$ line with a $\rho$ propagator lowers the degree of divergence by two unites. 
Indeed, if one focuses on the divergent part, the extra relative factor $g_\rho^2 f_\rho^2$ of the new diagram can only be compensated by a factor $1/\Lambda^2$,
where $\Lambda$ is the cutoff scale. Therefore, diagrams with loops of NG bosons alone (and no transverse gauge field $\rho$) carry the largest superficial degree
of divergence.  
If they entail a breaking of the EW symmetry, then their sum will vanish in the two-site model, 
since one can set $g_\rho =0$ in their evaluation and by the previous argument the electroweak symmetry is exact in this limit.
The original superficial degree of divergence is thus lowered.
In particular, 1PI contributions to EWSB observables will be finite in the $SO(5)/SO(4)$ theory (with both $\rho^L$ and $\rho^R$) for 
$a_\rho = 1/\sqrt{2}$~\footnote{Here and in the following we  use the notation $a_\rho = 1/\sqrt{2}$ as a shorthand for $a_\rhoL = a_\rhoR =1/\sqrt{2}$. 
Similarly, $g_\rho = 0$ must be always interpreted as $g_\rhoL = g_\rhoR = 0$.}
if they are at most logarithmically divergent in the general case.

This power counting argument was used in Ref.~\cite{Panico:2011pw} to conclude that the $S$ and $T$ parameters are finite in the  $a_\rho =1/\sqrt{2}$ limit. 
In the case of the $S$ parameter one can easily prove that for $g_\rho =0$ there is no local counterterm for 1PI divergent contributions
to the $\langle W_\mu^3 B_\nu \rangle$ Green function that can be constructed in the two-site 
model compatibly with the $SO(5) \times SO(5)_H$ symmetry, see Appendix~\ref{app:two-site}.
Local operators built by including powers of the spurion $g_\rho$ can be generated at the cutoff scale through loops where both the heavier states and the $\rho$ circulate.
By power counting these effects are finite at the 1-loop level, and  lead to a contribution to the $S$ parameter that is suppressed by an additional factor 
$(g_\rho f/\Lambda)^2 = (g_\rho/g_*)^2$ compared to the naive estimate. They are thus subleading and can be  neglected if $g_\rho \ll g_*$.
As discussed in Section~\ref{sec:matching}, our calculation confirms that the 1PI divergence 
(hence the $\beta$-function of $c_3^+$) vanishes for $a_\rho = 1/\sqrt{2}$. The $S$ parameter
is thus calculable in terms of the renormalized $g_\rho$ and $\alpha_2$, which absorb the divergences associated to subdiagrams.
Things work differently for the $T$ parameter, however. It turns out that while the 1PI divergence to the $\langle W^1 W^1 \rangle - \langle W^3 W^3 \rangle$ 
Green function vanishes according to the argument of  Ref.~\cite{Panico:2011pw}, the $\beta$-function of $c_T$ does not vanish for $a_\rho = 1/\sqrt{2}$ and
there is still a dependence on $c_T$ in the final result which enters through the cancellation of the subdivergences.
This can be seen as follows.

First of all, we notice that in the theory above $m_\rho$ it is 
possible to embed $O_T$ into the  $(SO(5)\times SO(5)_H)$-invariant operator
\begin{equation} \label{eq:OTsymm}
\left( \Tr\!\left[ (d_\mu + 2 \tilde d^L_\mu + 2 \tilde d^R_\mu ) \chi(\pi,\eta) \right] \right)^2 \ \longrightarrow \
\left( \Tr\!\left[ (d_\mu(\pi) - \bar\rho^L_\mu - \bar\rho^R_\mu ) \chi(\pi) \right]_{5\times 5} \right)^2\, ,
\end{equation}
where the expression after the arrow is obtained by going to the unitary gauge $\eta =0$.
The simplest way to show that this operator is $SO(5)\times SO(5)_H$ invariant 
is through the  link field $\bar U(\pi,\eta)$, see Appendix~\ref{app:two-site}.
By expanding the square in Eq~(\ref{eq:OTsymm}) one obtains a linear combination of $O_T$ and 
other operators of the Lagrangian (\ref{eq:Lrho}) with coefficients satisfying the relations
\begin{equation} \label{eq:SO5relations}
\beta_{1L} =\beta_{1R} = - \beta_{LR} = -2 c_{T} \, , \qquad \beta_{2L} = \beta_{2R} = c_{T}\, .
\end{equation}
These are the relations which must be imposed on the coefficients of the Lagrangian~(\ref{eq:Lrho}) in order to 
recover the larger $SO(5)\times SO(5)_H$ global symmetry at the level of terms quadratic in~$\chi$. 
This means that invariance under $SO(5) \times SO(5)_H$  does not force $c_T$ to vanish, but simply to become correlated with the coefficients of other 
operators in the Lagrangian.

But how a non-vanishing $c_T$ is compatible with the fact that no EWSB occurs in the two-site model for $g_\rho =0$ ?
In this limit,  there is an $[SU(2)_L \times U(1)_Y] \times SO(5)_H \to [SU(2)_L \times U(1)_Y]_d$ invariance after the EW gauging which gives 10 NG bosons. 
Four of these are eaten to give mass to the $W^a_\mu$ triplet and to the hypercharge,
while the others remain massless and transform as a $2_{1/2}$ (the composite Higgs doublet), and a $1_{\pm 1}$ of the
unbroken $[SU(2)_L \times U(1)_Y]_d$.~\footnote{One can also describe 
the same particle content in terms of the NG bosons of $SO(5)_H/[SU(2)_L \times U(1)_Y]$ plus massive spin-1 resonances ($W_\mu$ and $B_\mu$).}.
In particular, the unbroken global symmetry  forces the $W^i$ to form a degenerate triplet.
Compatibly with this,  the operator in Eq.~(\ref{eq:OTsymm}) does not lead to any splitting 
between $W^3$ and $W^{1,2}$:  the term $W_\mu^3 W^{3\, \mu}$  contained
in the expansion of $O_T$ is exactly canceled by a similar contribution from the other operators in the Lagrangian~(\ref{eq:Lrho}) as a consequence of the 
relations~(\ref{eq:SO5relations}). One has:
\begin{equation} \label{eq:counterterm}
\begin{split}
\Tr\!\left[ (d_\mu + 2 \tilde d^L_\mu + 2 \tilde d^R_\mu ) \chi \right] = & \,  \frac{g'}{f} \sqrt{2} \left( \partial^\mu \eta_\mu^{3L} \sin^2(\theta/2) +
\partial^\mu \eta_\mu^{3R} \cos^2(\theta/2)  - \frac{\sin\theta}{2}  \partial_\mu \pi^3 \right) \\
& + g' \left( B_\mu - \rho_\mu^{3L} \sin^2(\theta/2)  - \rho_\mu^{3R} \cos^2(\theta/2) \right) + \dots  
\end{split}
\end{equation}
Since no corresponding
counterterm is contained in  Eq.~(\ref{eq:OTsymm}), any 1PI contribution to the  Green function $\langle W^1 W^1 \rangle - \langle W^3 W^3  \rangle$ must be finite,
in agreement with the power counting argument of Ref.~\cite{Panico:2011pw}.
This is however not sufficient to conclude that the $T$ parameter is finite, since non-1PI diagrams also contribute and can be divergent.~\footnote{We thank
G.~Panico and A.~Wulzer for discussions on this point.} Our calculation in Appendix~\ref{sec:altmatchingcT}
 indeed shows that a divergent contribution arises from subdiagrams
through the 1-loop correction to the $\rho$ propagator. The associated counterterm is contained in the operator~(\ref{eq:OTsymm}), whose coefficient $c_T$ thus enters 
in the expression of the $T$ parameter. 

It is interesting to notice that the $T$ parameter
can also be extracted from the  Green function $\langle \pi^3 \pi^3 \rangle$,
as done in Section~\ref{sec:matching}, for which a 1PI divergent contribution does exist. The corresponding counterterm $(\pi^3)^2$  is 
contained in Eq.~(\ref{eq:OTsymm}), 
and it is not in clash with the argument of Ref.~\cite{Panico:2011pw}. This is because $\pi^3$ appears in the linear combination of NG bosons, the one in 
parenthesis in the first line of Eq.~(\ref{eq:counterterm}), that is eaten to give mass to the hypercharge for $g_\rho =0$.~\footnote{For $\theta =0$ the 
NG boson eaten  by the hypercharge is   $\eta^{3R}$, while the $\eta^{aL}$ are eaten to give mass to the $W$ triplet.}
The $\langle \pi^3 \pi^3 \rangle$ Green function thus does not break the $[SU(2)_L \times U(1)_Y]_d$ symmetry and can  be divergent.

Although it depends on $c_T$,  the $T$ parameter can still be regarded  as a calculable quantity in the two-site limit, up to $g_\rho^2/g_*^2$ effects.
This is because the operator (\ref{eq:OTsymm}) gives a custodially-breaking shift to the mass of the neutral $\rho$'s,
so that $c_T$ can be rewritten
in terms of the difference of charged and neutral renormalized $\rho$ masses. 
In this sense $T$, similarly to~$S$, is calculable in terms of parameters related to the $\rho$, which can be
fixed experimentally by measuring its properties.

\section{Electroweak parameters at 1 loop}
\label{sec:calculation}

\begin{figure}
\begin{center}
\includegraphics[width=0.30\textwidth]{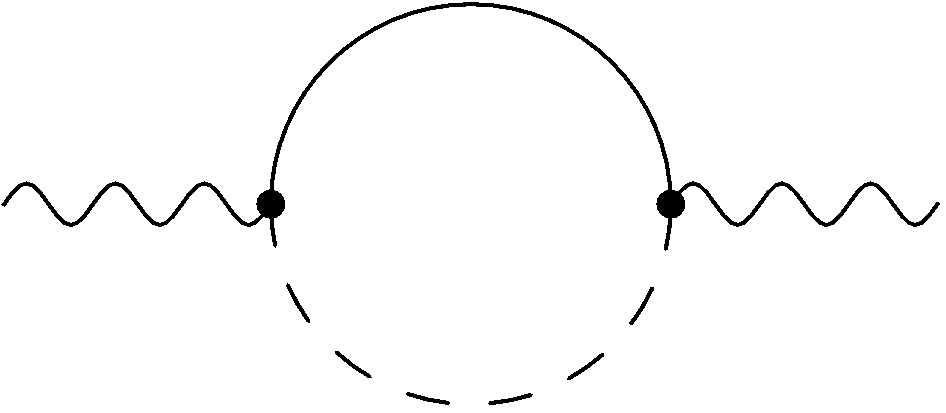}
\hspace{0.8cm}
\includegraphics[width=0.30\textwidth]{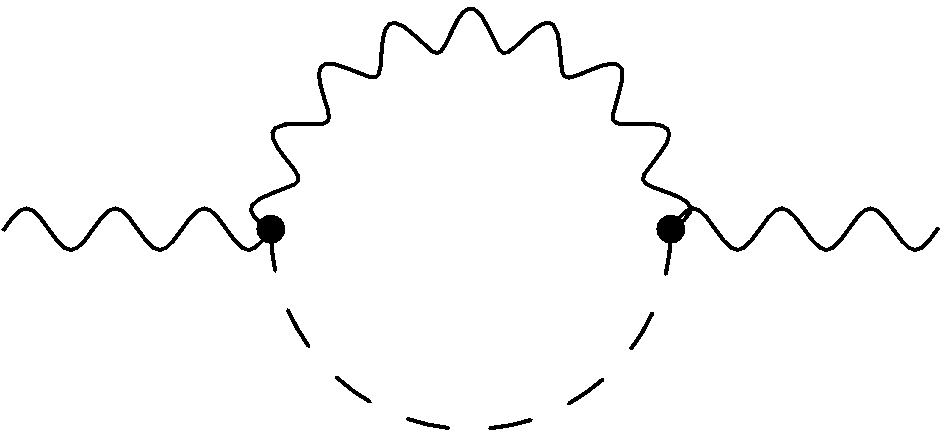}
\end{center}
\caption{\small
One-loop diagrams relative to the Higgs contribution to the epsilon parameters. Wavy, continuous and dashed lines denote respectively gauge fields ($W^\pm$ and $Z$),
NG bosons of $SO(4)/SO(3)$ ($\pi^{1,2,3}$) and the Higgs boson.}
\label{fig:Higgscontribution}
\end{figure}

Oblique corrections to the electroweak precision observables at the $Z$-pole are conveniently described by the three $\eps$ 
parameters~\cite{Altarelli:1990zd,Altarelli:1991fk}
\begin{equation}
\begin{split}
\eps_1 & = e_1 - e_5 \\
\eps_2 & = e_2 - s_W^2 e_4 - c_W^2 e_5 \\
\eps_3 & = e_3 + c_W^2 e_4 - c_W^2 e_5
\end{split}
\end{equation}
defined in terms of the following vector-boson self energies:
\begin{equation} \label{eq:selfenergies}
\begin{split}
e_{1} & =  \frac{1}{m_{W}^{2}}\left(A_{33}(0)-A_{W^{+}W^{-}}(0)\right),\\
e_{2} & =  F_{W^{+}W^{-}}(m_{W}^{2})-F_{33}(m_{Z}^{2}),\\
e_{3} & =  \frac{c_W}{s_W}F_{3B}(m_{Z}^{2}),\\
e_{4} & =  F_{\gamma\gamma}(0)-F_{\gamma\gamma}(m_{Z}^{2}),\\
e_{5} & =  m_{Z}^{2}F_{ZZ}^{\prime}(m_{Z}^{2})\, .
\end{split}
\end{equation}
Here $s_W$ ($c_W$) denotes the sine (cosine) of the Weinberg angle and, according to the standard notation, the vacuum polarizations are
decomposed as
\begin{equation}\label{eq:Pi}
\Pi_{ij}^{\mu\nu}(q) =  -i\eta^{\mu\nu} \left( A_{ij}(0) + q^2 F_{ij}(q^2) \right) + q^\mu q^\nu \text{ terms}\, .
\end{equation}
There are two kind of modifications to the self-energies~(\ref{eq:selfenergies}) from new physics in our model.
The first is due to the virtual exchange of the spin-1 resonances, 
which at energies $E \sim m_Z\ll m_\rho$ can be parametrized in terms  of local operators of the effective Lagrangian~(\ref{eq:effLpion}).
The tree-level contribution of these local operators to physical observables is a pure short-distance effect, while their insertion in 1-loop
diagrams with light fields contains also a long-distance part.
The second modification comes from the fact that the composite Higgs has non-standard couplings with the electroweak vector bosons.
The bulk of the correction in this case is given by a logarithmically divergent part that can be easily computed in the low-energy theory with
light fields~\cite{Barbieri:2007bh}.
Extracting the finite part instead requires fully recomputing the Higgs contribution to the vector boson self energies in Fig.~\ref{fig:Higgscontribution}, 
%
%
as pointed out in Ref.~\cite{Orgogozo:2012ct}.
Since the Higgs boson is light, this is a long-distance effect.
It is so even if  the compositeness scale is large, $f \gg v$,  so that the shifts of the Higgs couplings to vector bosons are parametrized by  local operators
at low energies. Indeed, the insertion of these local operators into the 1-loop diagrams of Fig.~\ref{fig:Higgscontribution} contains both long- and short-distance 
contributions.~\footnote{The divergent part of the diagrams corresponds to a renormalization
of the local operators of the effective Lagrangian, and it is thus a short-distance effect. The finite part is instead genuinely long distance.}

\begin{figure}
\begin{center}
\includegraphics[width=0.27\textwidth]{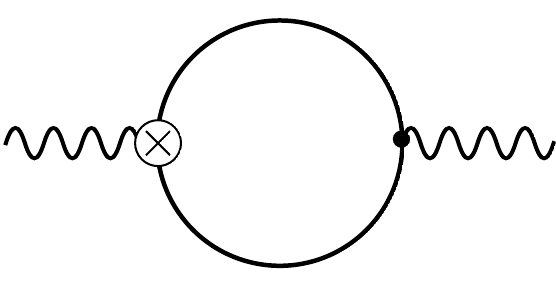}
\end{center}
\caption{\small
One-loop diagram with one insertion of $O_{3}^\pm$ and $O_4^\pm$ (crossed vertex) contributing to the running of $\tilde c_{2W}$ and $\tilde c_{2B}$ in the low-energy theory.
Wavy and continuous lines denote respectively gauge fields ($W$ and $B$) and NG bosons of $SO(5)/SO(4)$ ($\pi^{\hat a}$).}
\label{fig:cSinsertion}
\end{figure}

We have performed a calculation of the  $\eps_i$ at the 1-loop level including all the contributions mentioned above.
We have used dimensional regularization and performed a minimal subtraction of the divergences 
($\overline{MS}$ scheme). We choose to work in the Landau gauge for the elementary gauge fields, $\partial^\mu W_\mu^i = 0 = \partial^\mu B_\mu$, 
since it conveniently preserves the custodial invariance of the strong sector and leads to massless (hence degenerate) NG bosons $\pi^{1,2,3}$.
The one-loop contribution from the spin-1 resonances is computed through a matching procedure. 
We integrate out the~$\rho$ at a scale $\mu \sim m_\rho$ and match with a low-energy Lagrangian which has the same form of Eq.~(\ref{eq:effLpion}).
Its coefficients will be denoted by $\tilde c_i(\mu)$, where the tilde distinguishes them from the corresponding quantities in the full theory.
By working in such low-energy theory and defining the shifts to the
epsilons to be $\Delta\eps_i = \eps_i - \eps_i^{SM}$, we find 
\begin{align}
\label{eq:eps1}
\begin{split}
\Delta\epsilon_{1} =& \, -\frac{3g^{\prime2}}{32\pi^{2}}\sin^{2}\!\theta \left[\log\frac{\mu}{m_{Z}}+f_{1}\!\left(\frac{m_h^2}{m_Z^2}\right)\right]
                                  -2 g'^2 \sin^{2}\!\theta \, \tilde{c}_{T} \, , 
\end{split} \\[0.6cm]
\label{eq:eps2}
\begin{split}
\Delta\epsilon_{2} =& \, \frac{g^{2}}{192\pi^{2}}\sin^{2}\!\theta  \, f_{2}\!\left(\frac{m_h^2}{m_Z^2}\right) 
                                  + 2 m_W^2 g^2  \left(  \tilde c_{2W} \cos^4\frac{\theta}{2} + \tilde c_{2B} \sin^4\frac{\theta}{2} \right) \\[0.15cm]
   & \, + \frac{g^4}{24\pi^2} \sin^2\!\theta  \, \cos^4\frac{\theta}{2}  
       \left[ \left( \tilde c_3^+ + \tilde c_3^- \right) -\frac{1}{2} \left( \tilde c_4^+ + \tilde c_4^- \right) \right] \log\frac{\mu}{m_Z} \\[0.15cm]
   & \, + \frac{g^4}{24\pi^2} \sin^2\!\theta  \,  \sin^4\frac{\theta}{2} 
      \left[ \left( \tilde c_3^+ - \tilde c_3^- \right) - \frac{1}{2} \left( \tilde c_4^+ - \tilde c_4^- \right)  \right] \log\frac{\mu}{m_Z} \, ,
\end{split} \\[0.6cm]
\label{eq:eps3}
\begin{split}
\Delta\epsilon_{3}  =& \, \frac{g^{2}}{96\pi^{2}}\sin^{2}\!\theta \left[\log\frac{\mu}{m_{Z}}+f_{3}\!\left(\frac{m_h^2}{m_Z^2}\right)\right]
                                   -2g^{2}\sin^{2}\!\theta \, \tilde{c}_{3}^+ \, .
\end{split}
\end{align}
The first term in each equation corresponds to the 
Higgs contribution of Fig.~\ref{fig:Higgscontribution}~\footnote{It can be found from the Higgs 
contribution in the SM by considering that the Higgs couplings to vector bosons are rescaled by a factor $\cos\theta$, so that 
$\eps_i|_{Higgs} = \cos^2\!\theta\, \eps_i^{SM}|_{Higgs}$, hence  $\Delta \eps_i|_{Higgs} = - \sin^2\!\theta\, \eps_i^{SM}|_{Higgs}$.} and agrees with the results
of Ref.\cite{Orgogozo:2012ct}.
The explicit expression of the functions $f_{1,2,3}$  is given in Appendix~\ref{app:formulas}.
The coefficients $\tilde c_3^+, \tilde c_T, \tilde c_{2W}, \tilde c_{2B}$  encode the short-distance contribution from the $\rho$ and from cutoff states, 
and are in one-to-one correspondence with the parameters $S, T, W, Y$  defined in Refs.~\cite{Peskin:1991sw,Barbieri:2004qk}.
The latter are introduced through an expansion of the self energies~(\ref{eq:Pi}) in powers of~$q^2$ and 
parametrize the contribution from  new heavy physics. 
At the tree level one can  identify 
\begin{equation}
\label{eq:naiveid}
\begin{aligned} 
\hat S & = -2g^{2}\sin^{2}\!\theta \, \tilde{c}_{3}^+\, , \quad
& W & = -2 m_W^2 g^2 \left(  \tilde c_{2W} \cos^4\frac{\theta}{2} + \tilde c_{2B} \sin^4\frac{\theta}{2} \right) \\[0.3cm]
\hat T & = -2 g'^2 \sin^{2}\!\theta \, \tilde{c}_{T}\, , 
& Y & = -2 m_W^2 g^2 \left(  \tilde c_{2W} \sin^4\frac{\theta}{2} + \tilde c_{2B} \cos^4\frac{\theta}{2} \right) \, ,
\end{aligned}
\end{equation}
where $\hat S= (\alpha_{em}/4s_W^2) S$ and $\hat T =\alpha_{em} T$~\cite{Barbieri:2004qk}.
The naive estimate of $W$ and $Y$ is suppressed by a factor $g^2/g_\rho^2$ compared to that of $\hat S$ and~$\hat T$~\cite{Barbieri:2004qk}. 
We thus included their contribution (i.e. the contribution of $\tilde c_{2W}$ and $\tilde c_{2B}$) only in~$\eps_2$, where it gives the leading effect.
At the 1-loop level, the expression of $S, T, W, Y$ includes the $\log\mu$ terms of Eqs.~(\ref{eq:eps1})-(\ref{eq:eps3}). 
These arise from the short-distance, logarithmically divergent  part of the Higgs contribution,  and exactly compensate the dependence of the 
$\tilde c_i$ on $\mu$ to give an RG-invariant result.
The finite part of the Higgs contribution is a genuinely long-distance correction to the SM, and it is not encoded by $S, T, W, Y$, although it is captured
by the $\Delta\eps_i$. These latter are also independent of $\mu$, being observable quantities:  the variation of the $\tilde c_i(\mu)$
is canceled by the  logarithms in Eqs.~(\ref{eq:eps1})-(\ref{eq:eps3}).
We find that the evolution of the $\tilde c_i$ is described by the RG equations
\begin{equation}
\begin{aligned}
\mu \frac{d}{d\mu} \tilde c_3^+(\mu) & = \frac{1}{192\pi^2}\, , \qquad& 
\mu \frac{d}{d\mu} \tilde c_{2W}(\mu) & = - \frac{g^2}{48\pi^2} \frac{\sin^2\!\theta}{m_W^2} 
    \left[ (\tilde c_3^+ + \tilde c_3^-) -\frac{1}{2}(\tilde c_4^+ + \tilde c_4^-) \right]   \\[0.5cm]
\mu \frac{d}{d\mu} \tilde c_T(\mu) & = -\frac{3}{64\pi^2}\, , &
\mu \frac{d}{d\mu} \tilde c_{2B}(\mu) & = - \frac{g^2}{48\pi^2} \frac{\sin^2\!\theta}{m_W^2} 
    \left[ (\tilde c_3^+ - \tilde c_3^-) - \frac{1}{2} (\tilde c_4^+ - \tilde c_4^-) \right] \, .
\end{aligned}
\end{equation}
Notice that the $\beta$-function of $\tilde c_{2W}, \tilde c_{2B}$ is proportional to $\tilde c_3^\pm$ and $\tilde c_4^\pm$, 
since the running of these coefficients arises from the 1-loop 
insertion of the operators $O_3^\pm$ and $O_4^\pm$ defined in Eq.~(\ref{eq:CCWZbasis}),
see Fig.~\ref{fig:cSinsertion}. 
%
%
Since $\tilde c_3^\pm$ and $\tilde c_4^\pm$ are generated at tree level at the matching scale, they should be included at 1-loop in the calculation of $\eps_2$.
The last two terms in Eq.~(\ref{eq:eps2})  account for the divergent part of the diagram of Fig.~\ref{fig:cSinsertion}
and cancel the $\mu$ dependence due to the running of $\tilde c_{2W}, \tilde c_{2B}$. 
An additional finite contribution from of the 1-loop insertion of $O_{3}^\pm$ and $O_{4}^\pm$  has been omitted for simplicity.
It is subleading by a factor $\log{\mu/m_Z}$ and its computation would require evaluating
additional diagrams with gauge fields circulating in the loop.

\subsection{Matching}
\label{sec:matching}

The explicit contribution of the spin-1 resonances to the $\tilde c_i$ can be obtained by integrating them out
and matching to the low-energy Lagrangian. We perform this matching at the 1-loop level. This requires working out at the same time the 
renormalization of the Lagrangian for the $\rho$, in order to derive the RG evolution of its parameters. 
We considered two choices to fix the gauge invariance associated with the $\rho$ field and checked that
they both lead to the same result for physical quantities: the first is the unitary
gauge, where the $\rho$ is described by the Lagrangian~(\ref{eq:Lrho}); the second is the Landau gauge $\partial^\mu \rho_\mu^a =0$, obtained by introducing the
NG bosons $\eta$ as discussed in Section~\ref{sec:HLS}.
In the following we will report results for the unitary gauge, and collect formulas for the Landau gauge in Appendix~\ref{sec:spin1renorm}.
Particularly relevant for our calculation is the running of $g_\rho$ and $\alpha_2$, since these  parameters enter at tree level in the 
expression of the $\eps_i$. 
In the unitary gauge we find
\begin{align}
\label{eq:RGgrho}
\mu\frac{d}{d\mu} g_\rho(\mu) & \equiv \beta_{g_\rho} = \frac{g_\rho^3}{16\pi^2}  \, \frac{2 a_\rho^4-85}{12} \\[0.2cm]
\label{eq:RGalpha2}
\mu\frac{d}{d\mu} \alpha_2(\mu) & \equiv \beta_{\alpha_2} =  \frac{a_\rho^2 (1-a_\rho^2)}{96\pi^2} \, , 
\end{align}
for both $\rho_L$ and $\rho_R$ (there is no mixed renormalization of left and right parameters at the 1-loop level).
Other details on the renormalization of the $\rho$ Lagrangian can be found Appendix~\ref{sec:spin1renorm}.

A few remarks should be made about our calculation. First of all,
we will compute the Green functions relevant for the matching by neglecting the masses of the Higgs and of the vector bosons. 
This implies a relative error of order $m_h^2/m_\rho^2$, which is of the same size of the error due to the truncation of the effective Lagrangian
to the leading derivative operators (of $O(p^4)$ in the case of $\eps_{1,3}$ and $O(p^6)$ for $\eps_2$).
Infrared divergences are regulated by introducing a small common (hence custodially-preserving) mass $\lambda$ for the NG bosons. The dependence on 
$\lambda$ cancels out when matching the full and low-energy theories.
Second, the expressions for the $\tilde c_i$ reported in this section are 
obtained by including the contribution of $\alpha_{1,2}$ only at the tree level. 
This is justified if these coefficients are generated at the 1-loop level at the cutoff scale~$\Lambda$. The additional contribution from  $\alpha_{2}$ at 1-loop
is reported in Appendix~\ref{sec:alpha2oneloop}.
Finally, our formulas will include the contribution of both the $\rho_L$ and the $\rho_R$. In case only one resonance is present in the theory, 
$\tilde c_3^+$ and $\tilde c_T$ have the same expression for both $\rho_L$ and $\rho_R$, whereas $\rho_L$ only generates $\tilde c_{2W}$, and $\rho_R$ only $\tilde c_{2B}$.
This follows from a simple symmetry argument. Given a theory with a $\rho_L$, the case with a $\rho_R$ is obtained by performing a $P_{LR}$ transformation
on the strong dynamics. The equality of $\tilde c_3^+$ and $\tilde c_T$ then follows from the invariance of the operators $O_3^+$ and $O_T$ under such transformation.
On the other hand, acting with $P_{LR}$ interchanges $O_{2W}$ with $O_{2B}$, so that the expression of $\tilde c_{2W}$ in a theory with a $\rho_L$ equals that
of $\tilde c_{2B}$ in a theory with $\rho_R$. We report the corresponding expressions in Appendix~\ref{sec:singlerho} for convenience.

\begin{figure}
\begin{center}
\includegraphics[width=0.27\textwidth]{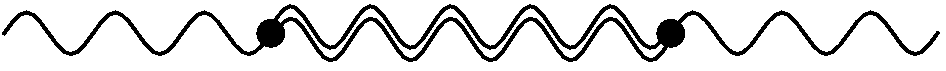}
\end{center}
\caption{\small
Tree-level diagram contributing to the $\langle W^3_\mu B_\nu \rangle$ Green function. Single and double wavy lines denote respectively the elementary
gauge fields ($W$ and $B$) and the $\rho$.}
\label{fig:S_tree}
\end{figure}

\begin{figure}
\vspace*{0.3cm}
\begin{center}
\includegraphics[width=0.27\textwidth]{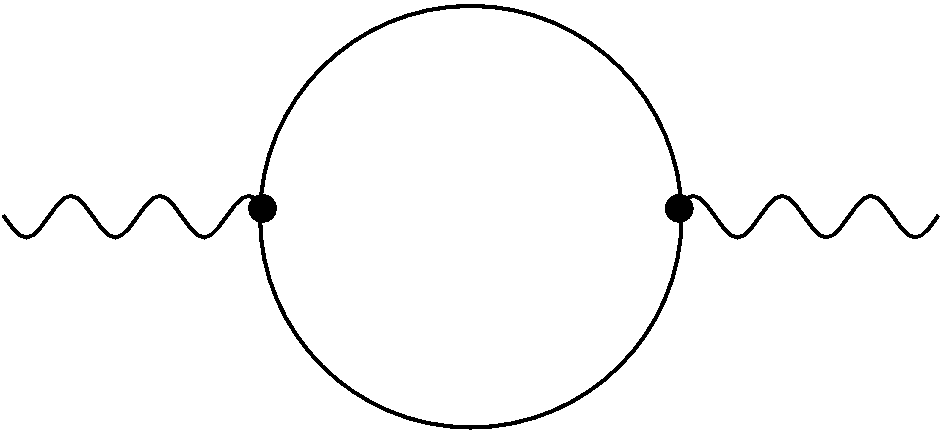}
\end{center}
\vspace*{-0.3cm}
\caption{\small
Diagram with a loop of NG bosons contributing to the $\langle W^3_\mu B_\nu \rangle$ Green function. Wavy and continuous lines denote respectively
the elementary gauge fields ($W$ and $B$) and the NG bosons ($\pi^{\hat a}$ and $\eta$).}
\label{fig:S_onlyNGbosons}
\end{figure}

\begin{figure}
\begin{center}
\includegraphics[width=0.28\textwidth]{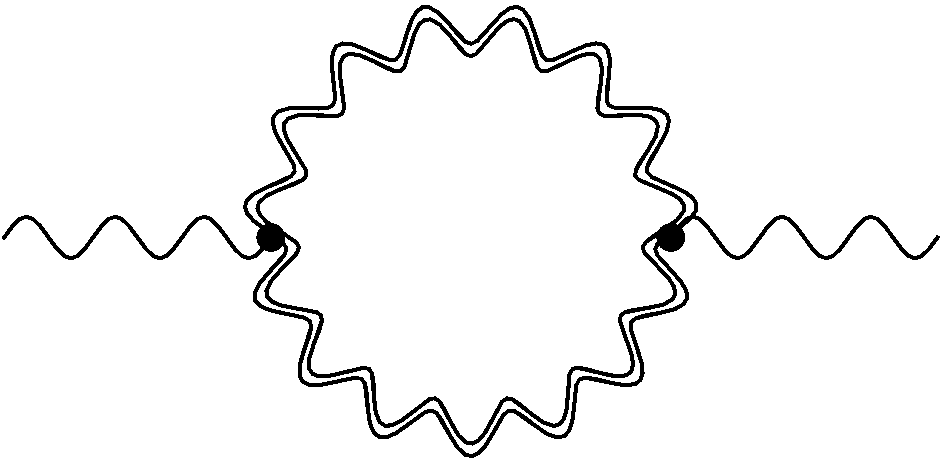}
\hspace{0.8cm}
\includegraphics[width=0.28\textwidth]{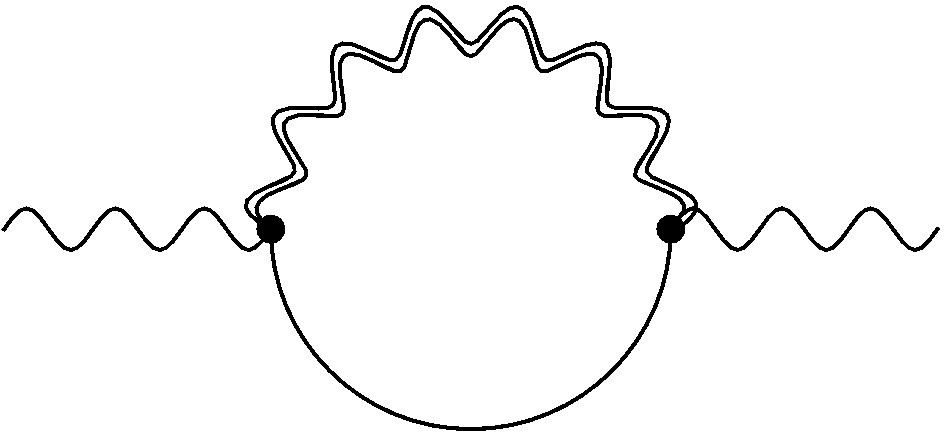}
\hspace{0.7cm}
\includegraphics[width=0.28\textwidth]{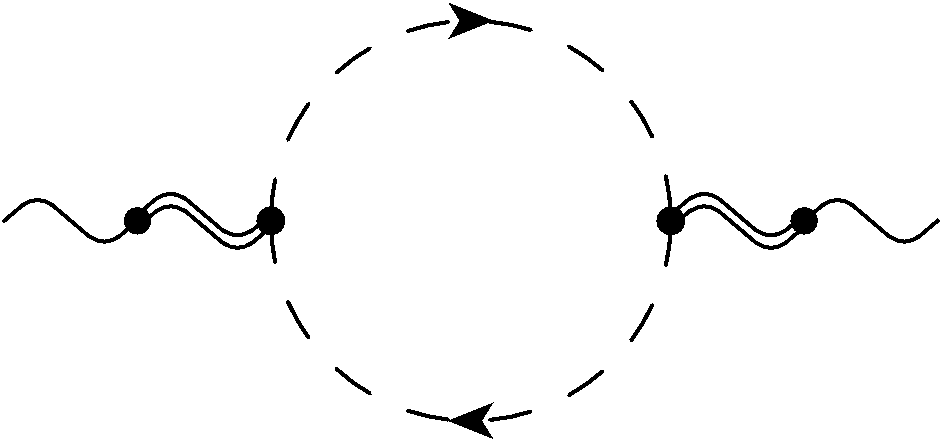}
\\[0.8cm]
\includegraphics[width=0.28\textwidth]{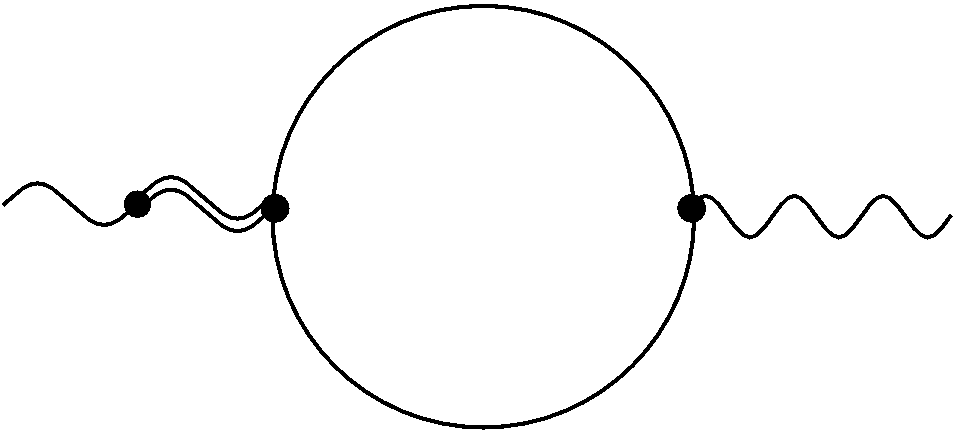}
\hspace{0.8cm}
\includegraphics[width=0.28\textwidth]{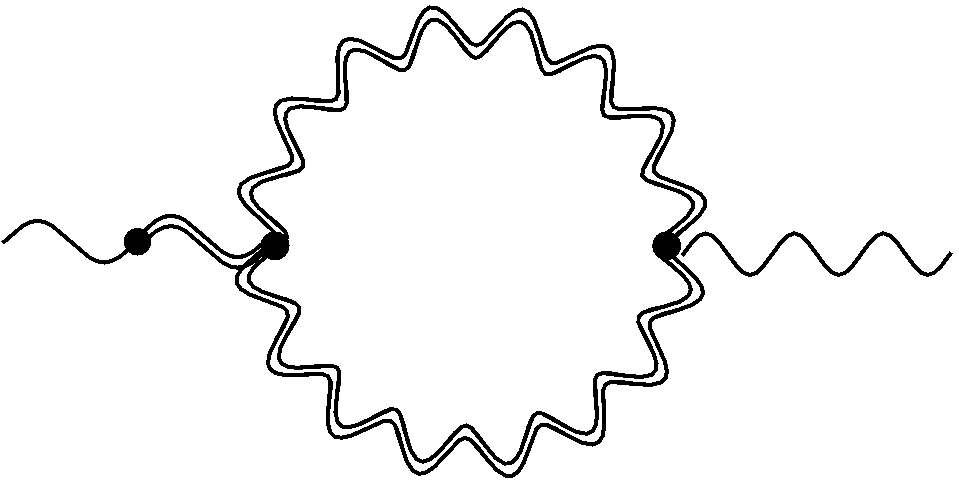}
\hspace{0.7cm}
\includegraphics[width=0.28\textwidth]{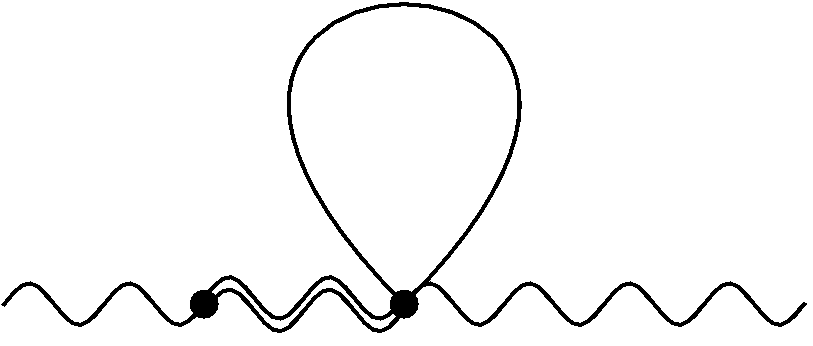}
\\[0.8cm]
\includegraphics[width=0.28\textwidth]{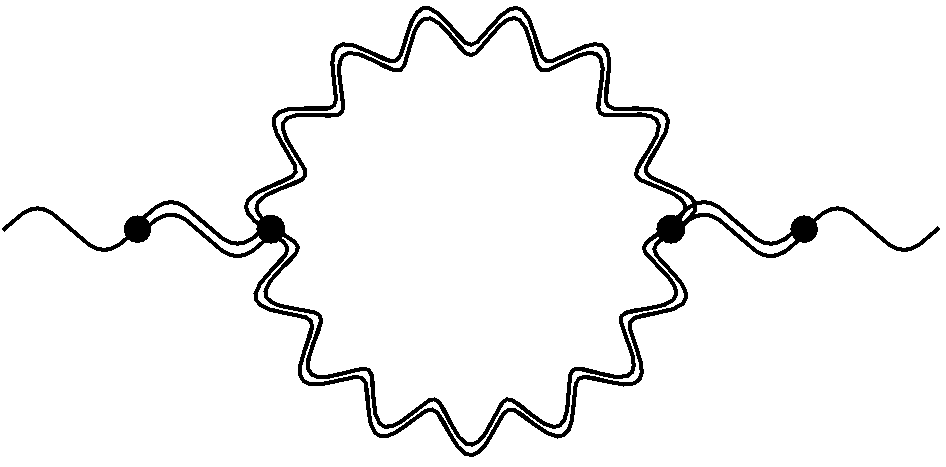}
\hspace{0.7cm}
\includegraphics[width=0.28\textwidth]{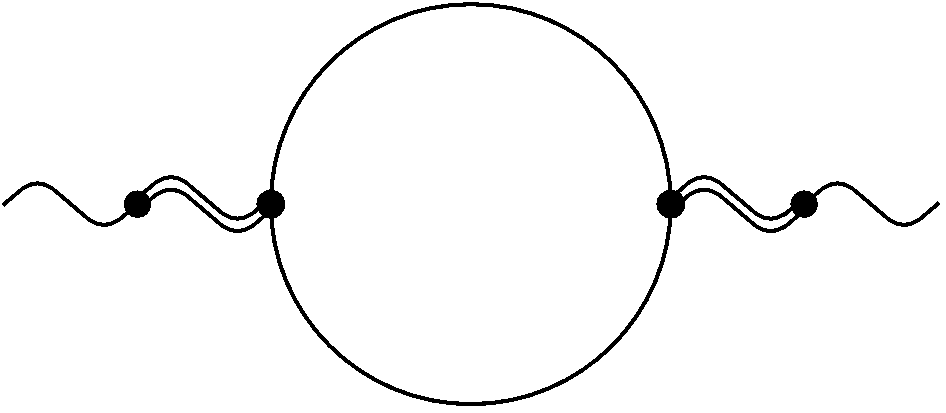}
\hspace{0.9cm}
\includegraphics[width=0.235\textwidth]{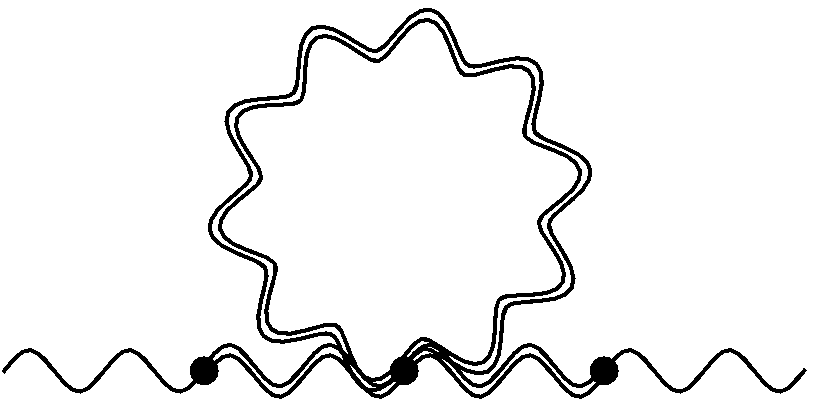}
\end{center}
\caption{\small
One-loop diagrams with $\rho$ exchange contributing to the $\langle W^3_\mu B_\nu \rangle$ Green function. 
Single and double wavy lines denote respectively the elementary gauge fields ($W$ and $B$) and the $\rho$; continuous and dashed
lines denote respectively the NG bosons ($\pi^{\hat a}$ and $\eta$) and the ghosts associated to the gauge fixing of the $\rho$ field.
The diagrams obtained by crossing those in the second line are not shown for simplicity.
}
\label{fig:S_loops}
\end{figure}

Let us start discussing the matching for $\tilde c_3^+$. We make use of the two-point Green function $\langle W^3_\mu B_\nu \rangle$,
in particular its derivative evaluated at $q^2=0$,  and match its expression in the full and effective theories.
We focus on the  leading contribution in $g^2$, thus considering  diagrams where only the $\rho$ and the NG bosons (i.e. no elementary
gauge field) circulate in the loop. These are the diagrams of Figs.~\ref{fig:S_tree},~\ref{fig:S_onlyNGbosons} and~\ref{fig:S_loops} for the full theory ($\rho$ + NG bosons) and 
of Fig.~\ref{fig:S_onlyNGbosons} for the effective theory (only NG bosons).
%
%
Neglecting diagrams with EW vector bosons circulating in the loop implies a relative error of order $g^2/g_\rho^2$.
Divergences from subdiagrams in the full theory are canceled by the addition of suitable counterterms.
The remaining divergence is associated with the running of $c_3^+$ between $m_\rho$ and $\Lambda$ due to loops of $\rho$'s and NG bosons.
We find 
\begin{equation} \label{eq:RGcS}
\mu\frac{d}{d\mu} c_3^+(\mu) \equiv \beta_{c_3^+} =  \frac{1}{192\pi^2} \left[  \frac{3}{2} + \frac{1}{4} a_{\rho_L}^2 (2 a_{\rho_L}^2 -7) 
                      +  \frac{1}{4} a_{\rho_R}^2 (2 a_{\rho_R}^2 -7)\right]  \, .
\end{equation}
Notice that $\beta_{c_3^+}$ (hence the associated divergence) vanishes for $a_{\rho_L} = a_{\rho_R} =1/\sqrt{2}$, 
in agreement with the symmetry argument of Section~\ref{sec:twosite}.
By matching the full and low-energy theories at a scale $\mu\sim m_\rho$, we obtain
\begin{equation} \label{eq:matchingcS}
\begin{split}
\tilde c_3^+(\mu) = & \, c_3^+(\mu)  -\frac{1}{2} \left( \frac{1}{4g_{\rho_L}^2} - \alpha_{2L} + \frac{1}{4g_{\rho_R}^2} - \alpha_{2R} \right) \\
 &  \, + \frac{1}{192\pi^2} \bigg[ \, \frac{3}{4} (a_{\rho_L}^2 + 28) \log\frac{\mu}{m_{\rho_L}} + \frac{3}{4} (a_{\rho_R}^2 + 28) \log\frac{\mu}{m_{\rho_R}} \\
 & \phantom{\, + \frac{1}{192\pi^2} \bigg[ \, } + 2 + \frac{41}{16} a_{\rho_L}^2  + \frac{41}{16} a_{\rho_R}^2 \bigg]\, .
\end{split}
\end{equation}
Obviously, since $\tilde c_3^+$ contributes to an observable such as $\Delta \eps_3$ (see Eq.~(\ref{eq:eps3})), its expression~(\ref{eq:matchingcS}) is the same in any gauge.
In fact, it turns out that even the $\beta$-function of $c_3^+$, Eq.~(\ref{eq:RGcS}), is gauge 
invariant at one loop.
The argument goes as follows. When working at the 1-loop level, the logarithms that appear in the expression of an observable determine the running of the
combination of the  parameters giving the tree-level contribution. Since the expression of the observable is gauge invariant, also the
RG evolution of such combination will be invariant. In the case of $\Delta \eps_3$, the tree-level contribution is given by the terms in the first line of Eq.~(\ref{eq:matchingcS}).
Furthermore, $(1/2g_\rho - \alpha_2 g_\rho)^2$ (for each $\rho$ species) also has a gauge 
invariant running, since it gives the tree-level  contribution to another observable: the pole residue of the $\rho$ two-point function~\cite{workinprogress}. 
Working in the approximation in which 1-loop effects from $\alpha_{1,2}$ are neglected, 
this in turn implies that  $(1/4g_\rho^2 - \alpha_2)$ has an invariant RG evolution,~\footnote{The running of the $\alpha_2^2$ term is of the same order of the neglected 
terms.}  hence the same follows for $c_3^+$. Clearly, when including the 1-loop contribution of $\alpha_2$ or going to two loops,  the running of $c_3^+$ acquires
a gauge-dependent part.

\begin{figure}
\begin{center}
\includegraphics[width=0.28\textwidth]{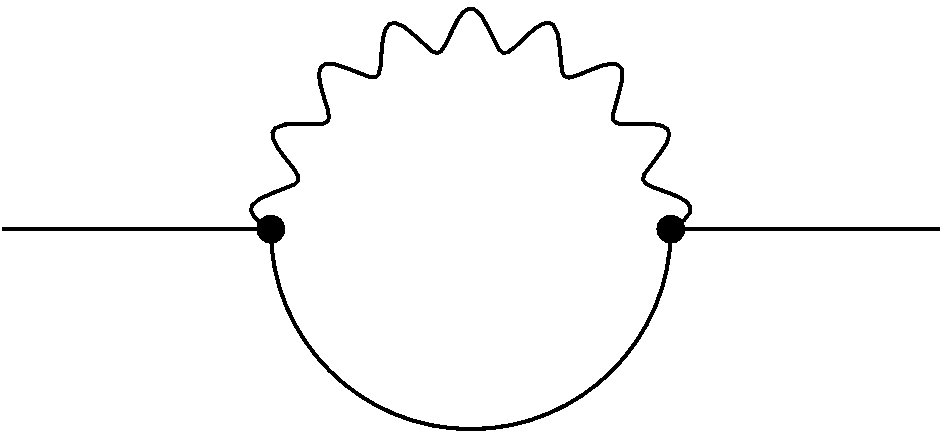}
\end{center}
\caption{\small
One-loop diagram with NG bosons contributing to the $\langle \pi^1 \pi^1 \rangle - \langle \pi^3 \pi^3 \rangle$ Green function. 
Wavy and continuous lines denote respectively the hypercharge gauge field $B$ and the NG bosons ($\pi^{\hat a}$ and $\eta$).
}
\label{fig:T_onlyNGbosons}
\end{figure}

\begin{figure}
\begin{center}
\includegraphics[width=0.28\textwidth]{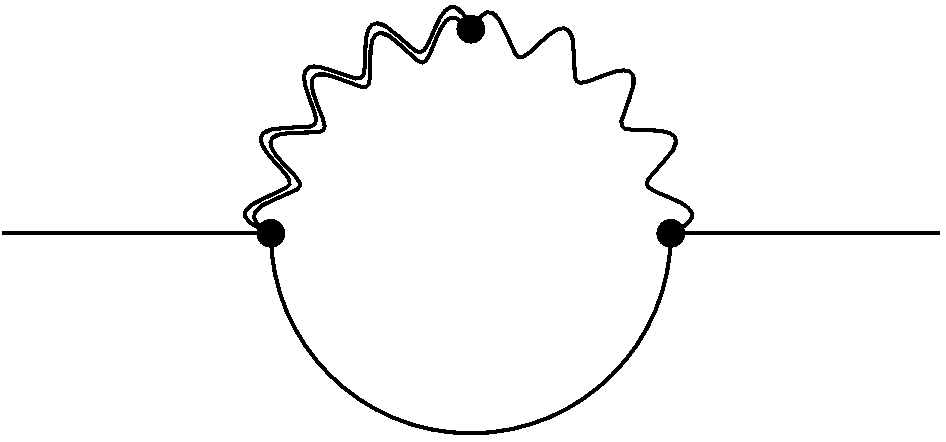}
\hspace{0.4cm}
\includegraphics[width=0.28\textwidth]{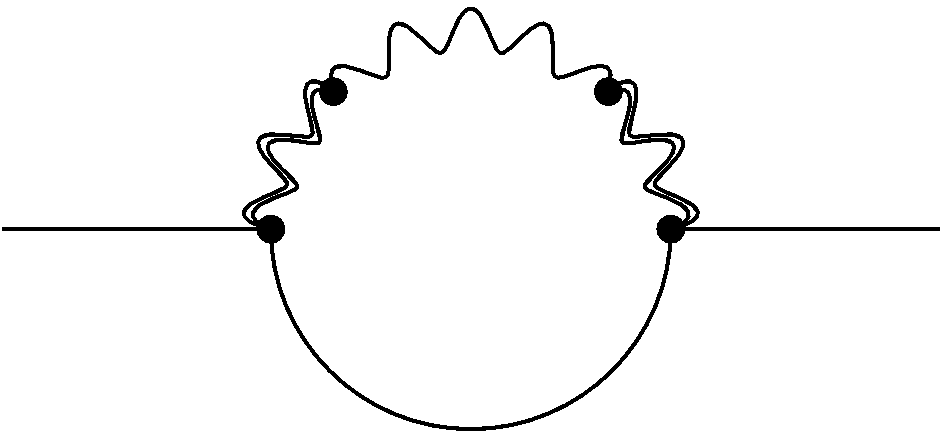}
\hspace{0.4cm}
\includegraphics[width=0.255\textwidth]{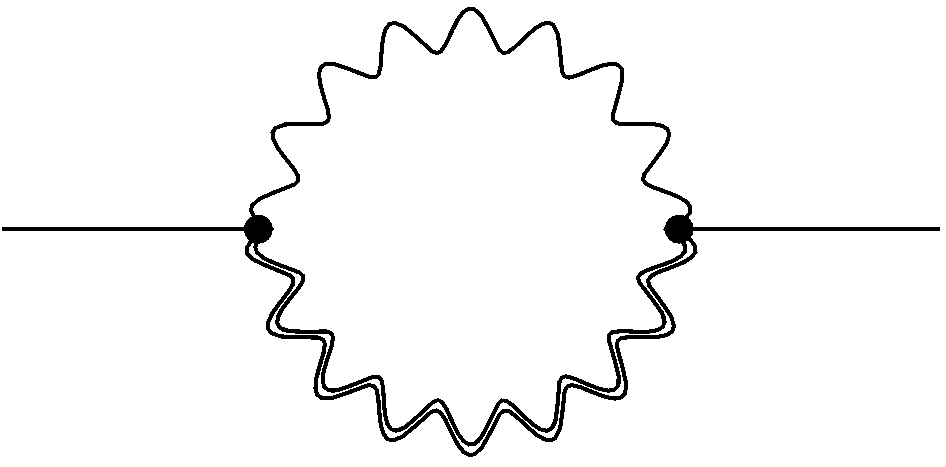}
\end{center}
\caption{\small
One-loop diagrams with $\rho$ exchange contributing to the $\langle \pi^1 \pi^1 \rangle - \langle \pi^3 \pi^3 \rangle$ Green function. 
Single and double wavy lines denote respectively the hypercharge gauge field $B$ and the $\rho$, while continuous 
lines denote the NG bosons ($\pi^{\hat a}$ and $\eta$).
The diagram obtained by crossing the first one is not shown for simplicity.
}
\label{fig:T_loops}
\end{figure}

Let us now turn to $\tilde c_T$. 
In order to extract it, we make use of the two-point Green function of the $\pi$ field, in particular we consider 
the custodially breaking combination $\langle \pi^1 \pi^1 \rangle - \langle \pi^3 \pi^3 \rangle$ and compute its derivative at $q^2 =0$.
This gives access to the coefficient of the operator $O_T$, as it follows from the expansion 
$\Tr[d_\mu \chi ] = g' \sin^2\!\theta \, (W^3_\mu - B_\mu) - g' \sin\theta \, (\partial_\mu \pi^3/f) + \dots$
In alternative, one can extract~$\tilde c_T$ by considering the combination $\langle W^1 W^1 \rangle - \langle W^3 W^3 \rangle$, as illustrated in 
Appendix~\ref{sec:altmatchingcT}.
The relevant 1-loop diagrams are shown in Figs.~\ref{fig:T_onlyNGbosons} and~\ref{fig:T_loops} for the full theory ($\rho$ + NG bosons), and in 
Fig.~\ref{fig:T_onlyNGbosons} for the low-energy theory of NG bosons.
%
%
Only diagrams where an elementary $B_\mu$ circulates contribute, as this latter gives the required breaking of custodial symmetry.
As for $\tilde c_3^+$, we neglect diagrams with further insertions of  EW vector bosons, since they are of higher order  in $g^2$.
The corresponding relative error is of order $g^2/g_\rho^2$.
Since there are no divergent subdiagrams,  the overall divergence in the full theory is associated with the running of $c_T$ between 
the scales $\Lambda$ and $m_\rho$.
We find: 
\begin{equation} \label{eq:RGcT}
\mu\frac{d}{d\mu} c_T(\mu) \equiv \beta_{c_T} = - \frac{3}{64\pi^2} \left(  1 - \frac{3}{4} a_{\rho_L}^2 - \frac{3}{4} a_{\rho_R}^2 + a_{\rho_L}^2 a_{\rho_R}^2 \right) \, .
\end{equation}
Since $c_T$ gives the only tree-level contribution to $\Delta\eps_1$ (see Eq.~(\ref{eq:matchingcT}) below), its RG evolution is gauge invariant.
One can see that $\beta_{c_T}$  does not vanish for $a_{\rho_L} = a_{\rho_R} =1/\sqrt{2}$.
This confirms the argument of Section~\ref{sec:twosite}, where it was noticed that 
a counterterm exists also in the $SO(5) \times SO(5)$ symmetric limit 
(see Eq.~(\ref{eq:OTsymm})), and no cancellation of the 1PI divergence of the $\langle \pi^1 \pi^1 \rangle - \langle \pi^3 \pi^3 \rangle$ Green function
is expected in this case. 
There is in fact a limit in which the  divergence partly cancels, as already discussed in Ref.~\cite{Orgogozo:2011kq} for a Higgsless model. 
Indeed, the diagram of Fig.~\ref{fig:T_onlyNGbosons} and the first two diagrams in Fig.~\ref{fig:T_loops} can be
combined into one where $B_\mu$ couples to the NG bosons through the effective vertex
\begin{center}
\includegraphics[width=0.5\textwidth]{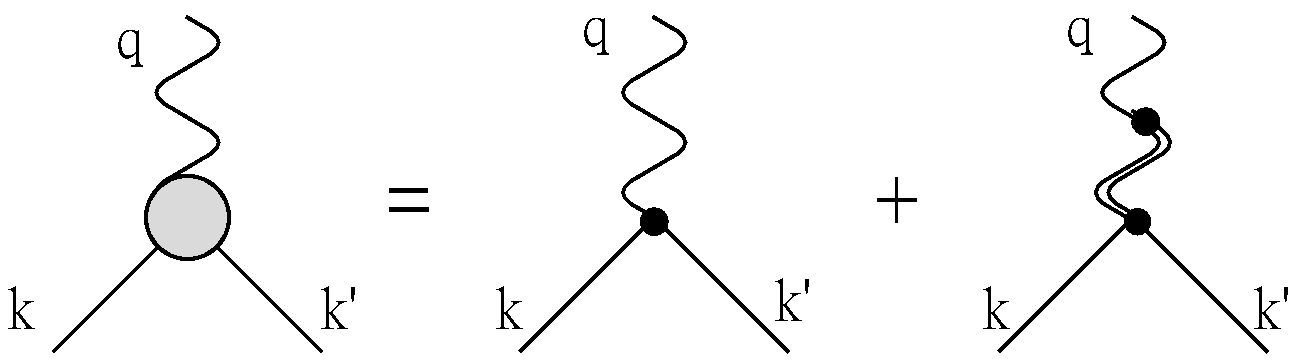}
\end{center}
where the $B\pi^{\hat a} \pi^{\hat b}$ form factor denoted by the gray blob is equal to
\begin{equation}
\begin{split}
\Bigg\{ 
 & \frac{1}{2} \bigg(1-a_{\rho_L}^{2}\sin^{2}\frac{\theta}{2}-a_{\rho_R}^{2}\cos^{2}\frac{\theta}{2}
   -\frac{a_{\rho_L}^{2} m_{\rho_L}^{2}\sin^{2}(\theta/2)}{q^{2}-m_{\rho_L}^{2}}-\frac{a_{\rho_R}^{2}m_{\rho_R}^{2}\cos^{2}(\theta/2)}{q^{2}-m_{\rho_R}^{2}}\bigg) \epsilon^{3\hat a\hat b}  \\
 & - \frac{1}{4} \bigg(\cos\theta + a_{\rho_L}^{2}\sin^{2}\frac{\theta}{2} - a_{\rho_R}^{2}\cos^{2}\frac{\theta}{2}
     +\frac{a_{\rho_L}^{2}m_{\rho_L}^{2}\sin^{2}(\theta/2)}{q^{2}-m_{\rho_L}^{2}} - \frac{a_{\rho_R}^{2}m_{\rho_R}^{2}\cos^{2}(\theta/2)}{q^{2}-m_{\rho_R}^{2}}\bigg) \\
 & \phantom{- \frac{1}{4}}\times \left( \delta^{\hat a 4} \delta^{\hat b 3} + \delta^{\hat a 3} \delta^{\hat b 4} \right) \Bigg\}  \left(k+k^{\prime}\right)^{\mu}+q^{\mu} \text{ terms}\, .
\end{split}
\end{equation}
In the limit $a_{\rho_L} = a_{\rho_R} = 1$ one obtains 
Vector Meson Dominance (VMD) for any value of~$\theta$, i.e. the form factor goes to 0 in the limit $q^2 \to \infty$.
Consequently, the diagram built with the effective vertex (i.e. the sum of the diagram in Fig.~\ref{fig:T_onlyNGbosons} and the first two of Fig.~\ref{fig:T_loops}) 
is finite. This does not imply, however, that the $\beta$-function of $c_T$ vanishes, since the last diagram of Fig.~\ref{fig:T_loops} is still divergent. One can 
explicitly check, indeed, that the coefficient of the logarithm in Eq.~(\ref{eq:RGcT}) does not vanish for $a_{\rho_L} = a_{\rho_R} = 1$.
By matching the full and low-energy theory at the scale $\mu$ we finally obtain
\begin{equation} \label{eq:matchingcT}
\begin{split}
\tilde c_T(\mu) = & \, c_T(\mu) - \frac{9}{256\pi^2} \bigg[ 
    a_{\rho_L}^2 \left( 1 - \frac{4}{3} a_{\rho_R}^2 \, \frac{m_{\rho_L}^2}{m_{\rho_L}^2- m_{\rho_R}^2} \right) \log\frac{\mu}{m_{\rho_L}} \\[0.15cm]
 & \phantom{\, c_T(\mu) - \frac{9}{256\pi^2} \bigg[ } 
    + a_{\rho_R}^2 \left( 1 - \frac{4}{3} a_{\rho_L}^2 \, \frac{m_{\rho_R}^2}{m_{\rho_R}^2- m_{\rho_L}^2} \right) \log\frac{\mu}{m_{\rho_R}}\\[0.15cm]
 & \phantom{\, c_T(\mu) - \frac{9}{256\pi^2} \bigg[ }+ \frac{3}{4} a_{\rho_L}^2 + \frac{3}{4} a_{\rho_R}^2 - \frac{5}{9}  a_{\rho_L}^2 a_{\rho_R}^2 \bigg]\, .
\end{split}
\end{equation}
Since $\tilde c_T$ contributes to the observable $\Delta\eps_1$, this expression is gauge invariant.

Finally, we discuss the matching to extract $\tilde c_{2W}$ and $\tilde c_{2B}$. We make use of the $\langle W_\mu W_\nu \rangle$ and $\langle B_\mu B_\nu \rangle$ 
Green functions, in particular we compute 
their second derivative evaluated at $q^2 =0$. Working at leading order in $g^2$, the diagrams in the full and effective theories are the same 
as in Figs.~\ref{fig:S_tree},~\ref{fig:S_onlyNGbosons} 
and~\ref{fig:S_loops}, where now the external gauge fields are either two $W$'s (to extract $\tilde c_{2W}$) or two $B$'s (to extract $\tilde c_{2B}$).
There is in fact one additional diagram, shown in Fig.~\ref{fig:cSinsertion},  which has to be included in the effective theory. It contains one insertion of the 
operators $O_3^\pm$ and $O_4^\pm$ defined in Eq.~(\ref{eq:CCWZbasis}).
As previously noticed, this contribution is relevant in the effective theory below $m_\rho$ since $\tilde c_{3}^\pm$ and $\tilde c_{4}^\pm$
are generated at the tree-level by the exchange of the~$\rho$.
Inserting $O_3^\pm$ and $O_4^\pm$ in a 1-loop diagram thus gives a contribution to $\tilde c_{2W}$ and $\tilde c_{2B}$ which is formally of the same order as that of 
the diagrams in Figs.~\ref{fig:S_tree}-\ref{fig:S_loops}.
In fact, such contribution is required in order to properly match the IR divergence of the full and low-energy theories. The cancellation occurs if  $\tilde c_3^\pm$ 
and $\tilde c_4^\pm$ 
are set to the value they have at tree-level for $\alpha_i = 0$ (that is: $\tilde c_3^\pm = -1/8g_{\rho_L}^2 \mp 1/8g_{\rho_R}^2$ and $\tilde c_4^\pm =0$)
when evaluating the diagram of  Fig.~\ref{fig:cSinsertion}; we will thus adopt this choice.~\footnote{When including the contribution of $\alpha_2$ at the 1-loop 
level,  as done in Appendix~\ref{sec:alpha2oneloop}, one should instead set $\tilde c_3^\pm = (-1/4g_{\rho_L}^2 + \alpha_{2L})/2 \pm (-1/4g_{\rho_R}^2+\alpha_{2R})/2$,
while including  $\alpha_1$ at the 1-loop level requires setting $\tilde c_4^\pm = (\alpha_{1L} \pm \alpha_{1R})/2$.}
There are no divergences left after removing those from subdiagrams through the renormalization of the $\rho$ mass and kinetic terms.
This implies that the running of the coefficients $c_{2W}$ and $c_{2B}$ vanishes in the full theory between $m_\rho$ and $\Lambda$:
\begin{equation}
\label{eq:RGc3W}
\mu \frac{d}{d\mu}c_{2W}(\mu) \equiv \beta_{c_{2W}} = 0\, , \qquad\qquad
\mu \frac{d}{d\mu}c_{2B}(\mu) \equiv \beta_{c_{2B}} = 0  \, .
\end{equation}
This result is independent of the choice of gauge.
Indeed,  by matching the full and low-energy theories we  obtain
\begin{align}
\label{eq:matchingc3W}
\begin{split}
\tilde c_{2W}(\mu) = & \, c_{2W}(\mu) - \frac{1}{2g_{\rho_L}^2 m_{\rho_L}^2} (1 - 2 \alpha_{2L} g_{\rho_L}^2)^2 \\
 & \, + \frac{1}{96\pi^2 m_{\rho_L}^2} \left[ 
  77 \log\frac{\mu}{m_{\rho_L}} + \frac{46}{5} - \frac{27}{32} a_{\rho_L}^2  \frac{\sin^2\!\theta}{1+\cos^2\!\theta} \left( 1+ \frac{g_{\rho_L}^2}{g_{\rho_R}^2} \right) \right] 
\end{split}
\\[0.5cm]
\label{eq:matchingc3B}
\begin{split}
\tilde c_{2B}(\mu) = & \, c_{2B}(\mu) - \frac{1}{2g_{\rho_R}^2 m_{\rho_R}^2} (1 - 2 \alpha_{2R} g_{\rho_R}^2)^2 \\
 & \, + \frac{1}{96\pi^2 m_{\rho_R}^2} \left[ 
  77 \log\frac{\mu}{m_{\rho_R}} + \frac{46}{5} - \frac{27}{32} a_{\rho_R}^2  \frac{\sin^2\!\theta}{1+\cos^2\!\theta} \left( 1+ \frac{g_{\rho_R}^2}{g_{\rho_L}^2} \right) \right] \, .
\end{split}
\end{align}
The tree-level contribution to $\Delta\eps_2$ comes from the combination of terms in the first line of the above equations. We already noticed  that
$(1/g_\rho - 2 \alpha_{2} g_\rho)^2$ has an invariant RG evolution at the 1-loop level; the same holds true  for $m_\rho$, since it gives the tree-level contribution 
to the pole mass. It thus follows that the RG evolution of $c_{2W}$ and $c_{2B}$ is also gauge invariant at one loop.

\section{Fit to the EW observables}
\label{sec:fit}

The results of the previous section can be used to perform a fit to the $\eps_i$.
It is convenient to express the corrections
$\Delta\eps_i$ in terms of  the parameters $g_\rho$, $\alpha_2$ and $m_\rho$ evaluated at the 
physical mass scale of the resonances $m_\rho^\text{pole}$.~\footnote{
For this evaluation we approximate $m_\rho^\text{pole}\simeq m_\rho$, the difference being of higher order.
}
This removes all the  logarithms originating from subdivergences leaving only those associated with the running of 
$O_3^+$, $O_T$, $O_{2W}$ and $O_{2B}$.
We will consider two benchmark scenarios: in the first (Scenario 1)  both $\rho_L$ and $\rho_R$ are present with equal masses and couplings (as 
implied for example by $P_{LR}$ invariance); in the second (Scenario 2) only a $\rho_L$ is included.
In either case the $\Delta\eps_i$ can be written as
\begin{align}
\label{eq:deps1final}
\Delta \epsilon_1 = &  \, - 2 g'^2 \sin^{2}\!\theta  \, c_T(\Lambda) - \frac{3g'^{2}}{32\pi^{2}}\sin^{2}\!\theta \left[ f_{1}\!\left(\frac{m_h^2}{m_Z^2}\right) + \log\frac{m_\rho}{m_{Z}}
                                    + \beta_{1} \log\frac{\Lambda}{m_\rho}  + \zeta_{1} \right]
\\[0.7cm]
\label{eq:deps2final}
\begin{split}
\Delta \epsilon_2 = & \, 2m_W^2 g^2  \left( c_{2W}(\Lambda) \cos^4\frac{\theta}{2}  + c_{2B}(\Lambda) \sin^4\frac{\theta}{2}  \right)
                                    - \gamma_2 \frac{g^2}{g_\rho^2} \frac{m_W^2}{m_\rho^2}  \left( 1 - 2 \alpha_2 g_\rho^2 \right)^2 \\
                                &  + \frac{g^2}{192\pi^2} \sin^{2}\!\theta \left[ f_{2}\!\left(\frac{m_h^2}{m_Z^2}\right) + \tilde\beta_2 \frac{g^2}{g_\rho^2} \log\frac{m_\rho}{m_{Z}}
                                     + \beta_{2} \frac{g^2}{g_\rho^2}  \log\frac{\Lambda}{m_\rho}  + \zeta_{2}  \frac{g^2}{g_\rho^2}  \right] 
\end{split} \\[0.7cm]
\label{eq:deps3final}
\begin{split}
\Delta \epsilon_3 = & \, - 2g^2 \sin^{2}\!\theta  \, c_3^+(\Lambda) + \gamma_3 \frac{g^2}{g_\rho^2} \sin^{2}\!\theta \left( 1 - 4 \alpha_2 g_\rho^2 \right) \\
                                & + \frac{g^{2}}{96\pi^{2}}\sin^{2}\!\theta \left[ f_{3}\!\left(\frac{m_h^2}{m_Z^2}\right) + \log\frac{m_\rho}{m_{Z}}
                                    + \beta_{3} \log\frac{\Lambda}{m_\rho}  + \zeta_{3}   \right]\, ,
\end{split}
\end{align}
where $g_\rho$, $\alpha_2$ and $m_\rho$ are evaluated at $\mu = m_\rho$ and the $O(1)$ coefficients $\beta_i$, $\tilde\beta_i$, $\zeta_i$, $\gamma_i$ are  reported 
in Table~\ref{tab:coeff} in the simplified limit  where 1-loop contributions from $\alpha_{1,2}$ are neglected.
%
\begin{table}[t]
\centering 
\begin{math}
\begin{array}{>{\displaystyle}c|@{\hspace{2em}}>{\displaystyle}c@{\hspace{2em}}>{\displaystyle} c}
 & \text{Scenario 1 ($\rho_L$ + $\rho_R$)} & \text{Scenario 2 ($\rho_L$)} \\
\hline && \\[-0.45cm]
\beta_1 & 1 - \frac{3}{2} a_\rho^2 + a_\rho^4  &  1 - \frac{3}{4} a_\rho^2 \\[0.3cm]
\zeta_1 & - \frac{9}{8} a_\rho^2 -\frac{1}{12} a_\rho^4 & -\frac{9}{16} a_\rho^2  \\[0.3cm]
\hline && \\[-0.5cm]
\beta_2 & 0 &  0 \\[0.3cm]
\tilde \beta_2 & -( 1+ \cos^2\!\theta) & -2 \cos^4\frac{\theta}{2} \\[0.3cm]
\zeta_2 &  ( 1+ \cos^2\!\theta) \left( \frac{23}{5a_\rho^2}  - \frac{27}{32}  \tan^2\frac{\theta}{2} \right)
             & \cos^4\frac{\theta}{2} \left( \frac{46}{5a_\rho^2}  - \frac{27}{32}  \tan^2\frac{\theta}{2}  \right)  \\[0.4cm]
\gamma_2 & \frac{1}{2} ( 1+ \cos^2\!\theta) & \cos^4\frac{\theta}{2}  \\[0.4cm]
\hline && \\[-0.45cm]
\beta_3 & \frac{3}{2} + \frac{a_\rho^2}{2} (2 a_\rho^2-7)  & \frac{5}{4} + \frac{a_\rho^2}{4}  (2 a_\rho^2-7) \\[0.3cm]
\zeta_3 & -2- \frac{41}{8} a_\rho^2 & -1 - \frac{41}{16} a_\rho^2 \\[0.4cm]
\gamma_3 & \frac{1}{2} & \frac{1}{4} 
\end{array}
\end{math}
\caption{\small 
Expression of the coefficients $\beta_i$, $\zeta_i$ and $\gamma_i$, defined in Eqs.~(\ref{eq:deps1final})-(\ref{eq:deps3final}), in the 
limit  where 1-loop contributions from $\alpha_{1,2}$ are neglected. Scenarios 1 includes $\rho_L$ and $\rho_R$ with equal masses and couplings,
while only $\rho_L$ is included in Scenario 2.
}
\label{tab:coeff}
\end{table}

Let us analyze Eqs.~(\ref{eq:deps1final})-(\ref{eq:deps3final}) and discuss the various terms. For each $\Delta\eps_i$  one can identify: a tree-level contribution from 
the exchange of spin-1 resonances (second term of Eqs.(\ref{eq:deps2final}) and (\ref{eq:deps3final})); a threshold correction due to Higgs compositeness (first term in 
square parenthesis); the IR running from $m_\rho$ down to $m_Z$, controlled by the low-energy $\beta$-function (second term in square parenthesis); the running from the 
cutoff $\Lambda$ to $m_\rho$, computed including the spin-1 resonances (third term in square parenthesis); a finite part from the 1-loop $\rho$ exchange
 (last term in square parenthesis). Finally, each  
$\Delta\eps_i$ receives a short-distance correction from physics at the cutoff scale, encoded by the coefficients $c_i(\Lambda)$ (first term in 
Eqs.(\ref{eq:deps1final})-(\ref{eq:deps3final})).

In the case of~$\eps_3$, the leading corrections  come from the tree-level contribution (of order $m_W^2/m_\rho^2$) and  the IR running.
Compared to the former, the latter effect is suppressed by a factor $g_\rho^2/16\pi^2$ but enhanced by $\log(m_\rho/m_Z)$. 
The 1-loop $\rho$ contribution is subleading because also suppressed by $g_\rho^2/16\pi^2$ and enhanced by the smaller logarithm 
associated with the running between $\Lambda$ and $m_\rho$. 
The contribution from cutoff physics encoded by $c_3^+(\Lambda)$ can be estimated through Naive Dimensional Analysis (NDA)~\cite{Manohar:1983md}.
If the dynamics at the scale~$\Lambda$ is maximally strongly coupled one expects $c_3^+(\Lambda) \sim 1/16\pi^2$, which leads to a correction of the same size 
of the finite part and thus subleading compared to the 1-loop $\rho$ contribution by a factor $\log(\Lambda/m_\rho)$. Although this logarithm is not large, 
since one does not expect a very large separation of scales,  it gives a parametric justification for including  the 1-loop effect of the $\rho$.
In general, if the cutoff dynamics is characterized by a coupling strength $g_*$, one naively expects $c_3^+(\Lambda) \sim 1/g_*^2$. For $g_\rho < g_* < 4\pi$
this implies a correction larger than the 1-loop $\rho$ contribution, though smaller than the tree-level one.
Interestingly, in the two-site limit (Scenario 1 with $a_{\rho} = 1/\sqrt{2}$) the  $SO(5)\times SO(5)_H$ global invariance of 
the theory below the cutoff  ensures $c_3^+(\Lambda) =0$, since the corresponding operator vanishes. Notice that $\beta_{c_3^+}$  vanishes also
in Scenario 2 for $a_{\rho_L} = 1$, although in that case no larger symmetry is realized that can enforce $c_3^+ (\Lambda) =0$. 
Similarly, no symmetry protection follows from the vanishing of $\beta_{c_T}$, $\beta_{c_{2W}}$ and  $\beta_{c_{2B}}$ for specific values of the parameters.

Similar estimates of the various terms hold for $\Delta\eps_1$, except there is no tree-level correction due to custodial invariance, 
so that the largest effect comes from the IR running. In the case of $\Delta\eps_2$, the contribution from the $\rho$ exchange (both at tree and loop level) is 
suppressed by a factor $(g^2/g_\rho^2)$ compared to the one entering $\Delta\eps_1$ and $\Delta\eps_3$. This is because
the leading short-distance contribution in the low-energy theory arises at $O(p^6)$ through the
operators $O_{2W}$, $O_{2B}$~\cite{Giudice:2007fh}. The RG evolution of these latter 
in turn proceeds through the 1-loop insertion of $O(p^4)$ operators, as discussed in the previous section, implying that the IR running contribution to $\Delta\eps_2$
is also suppressed by a factor $(g^2/g_\rho^2)$.  The only unsuppressed effect is the finite term from Higgs compositeness, which is however numerically small.
The overall shift to $\eps_2$ thus tends to be small and plays a minor role in the fit.

Besides the direct contributions to the $\Delta\eps_i$ described above there is also an indirect one from the evolution of $g_\rho$,
$m_\rho$ and $\alpha_2$ from the cutoff $\Lambda$ down to the scale $m_\rho$.
This is a numerically large effect if the $\Delta\eps_i$ are expressed in terms
of the values of these parameters at the scale $\Lambda$. The running of $g_\rho$, in particular, proceeds through a sizable and negative (for $a_\rho$ not too large) 
$\beta$-function, growing
quickly in the IR. This implies that for moderately large values of $g_\rho$ at the cutoff scale,  the gap $\Lambda/m_\rho$ cannot be too large otherwise $g_\rho$
would hit a Landau pole for $\mu > m_\rho$. For example,  $g_\rho(\Lambda) = 3$ gives a Landau pole at $\mu \simeq \Lambda/3.6$ in the unitary gauge.
Although the evolution of $g_\rho$ is gauge dependent, it gives a rough indication on how strongly coupled the theory of spin-1 resonances is. A more refined
estimate could make use for example of the combination $\lambda \equiv (1/g_\rho - 2 \alpha_2 g_\rho)^{-1}$ with gauge-invariant running.
Notice also that  $\beta_{g_\rho}$ will in general receive contributions also from other resonances lighter than the cutoff, like for example
the top partners, which could slow down the growth of $g_\rho$ in the IR and allow larger gaps.

In the following we  analyze the constraints from the current  electroweak data by constructing a $\chi^2$ function using the fit  of 
Refs.~\cite{Ciuchini:2013pca,Ciuchini:2014dea} to the $\Delta\eps_{i}$ and their theoretical predictions in Eqs.~(\ref{eq:deps1final})-(\ref{eq:deps3final}).~\footnote{We 
perform a 3-parameters fit by using Table~4 of Ref.~\cite{Ciuchini:2014dea} fixing $\eps_b = \eps_b^{SM}$. We derive the limits by determining the 
isocurves of $\Delta\chi^2$ corresponding to 3 degrees of freedom. Considering that $\eps_2$ does not vary much in our model (the new physics corrections is small),
one could adopt a more conservative choice and derive the isocurves with 2 degrees of freedom. This would lead to slightly stronger constraints, without qualitatively
affecting our conclusions.}
These latter will be evaluated 
in terms of the values of the parameters $g_\rho$, $m_\rho$ and $f$ at the scale $\mu =m_\rho$. In particular we use the identity $g_\rho = m_\rho/(a_\rho f)$ 
(Eq.~(\ref{eq:arho})) to rewrite $g_\rho$ in terms of $a_\rho$ and fix
\begin{equation}
\label{eq:xidef}
f(m_\rho) =  \frac{v}{\sqrt{\xi}}\, ,
\end{equation}
where $\xi \equiv \sin^2\!\theta$ and $v = 246\,$GeV is the electroweak scale.
This relation follows from the minimization of the Higgs potential generated by loops of heavy resonances.~\footnote{If electroweak symmetry breaking is triggered 
by the contribution
of a lighter set of resonances with mass $m_\Psi$, for instance the top partners, the relation becomes $f(m_\Psi) = v/\sqrt{\xi}$. In this case $f(m_\rho)$ can 
be derived by running from $m_\Psi$. Notice that $\beta_f$ is gauge invariant at one loop, since $f$ gives the tree-level correction to physical observables like
the on-shell  $\pi\pi \to \pi\pi$ scattering amplitude and the $W$ mass.}
The value of the remaining parameters $c_3^+$, $c_T$, $c_{2W}$, $c_{2B}$ is set to vanish at the scale $\Lambda$.
For the case of $c_3^+$, whose $\beta$-functions is gauge dependent when including the contribution from $\alpha_{1,2}$
at one loop, this condition is imposed in the unitary gauge.~\footnote{Equivalently,
one can fix $c_3^+(m_\rho)$ so that $c_3^+$ vanishes at $\mu =\Lambda$ in the unitary gauge.
The condition formulated  in this way at $\mu =m_\rho$ is gauge independent.}

\begin{figure}[h]
\begin{center}
\includegraphics[width=0.48\textwidth]{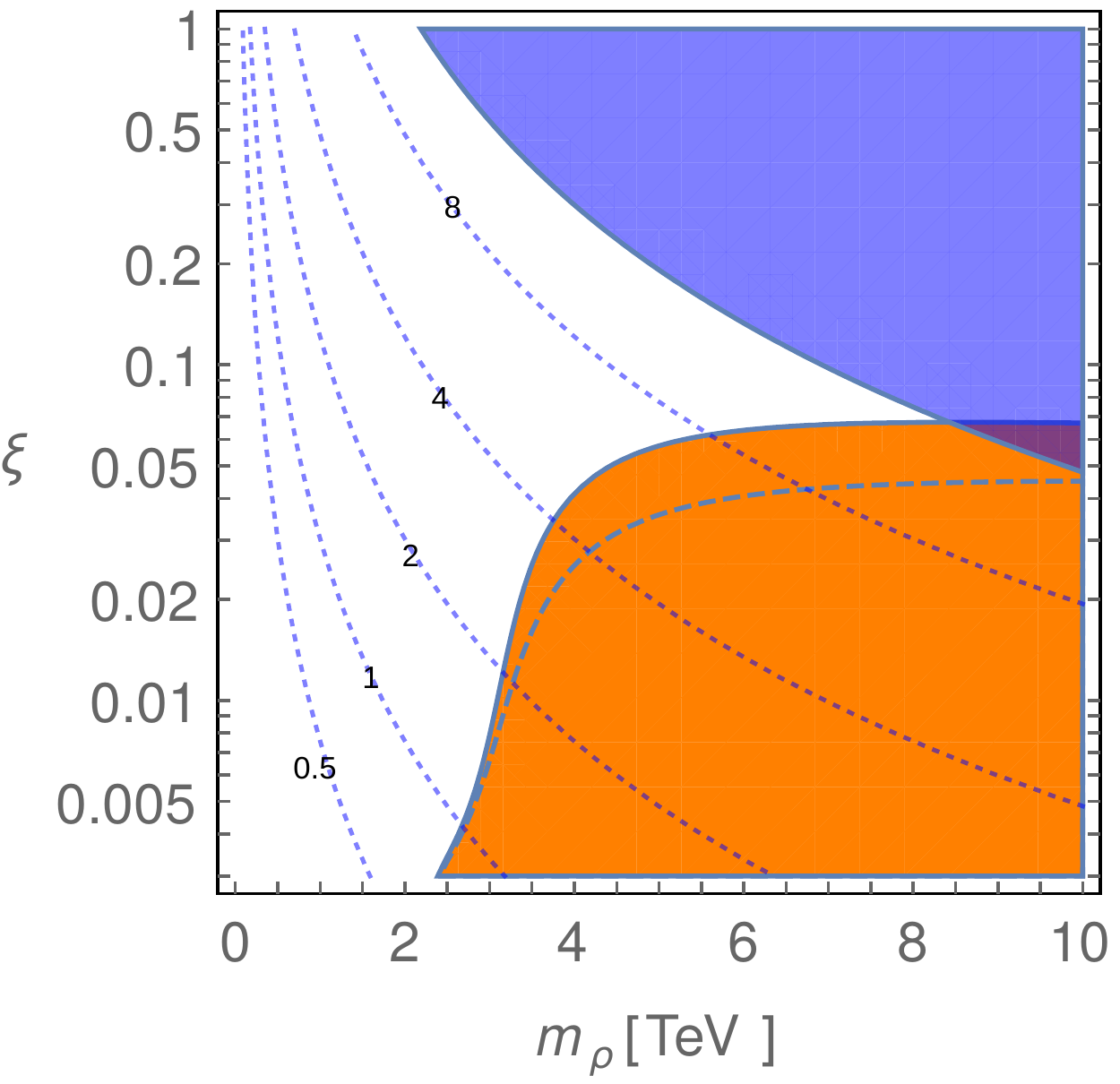}
\hspace{0.4cm}
\includegraphics[width=0.48\textwidth]{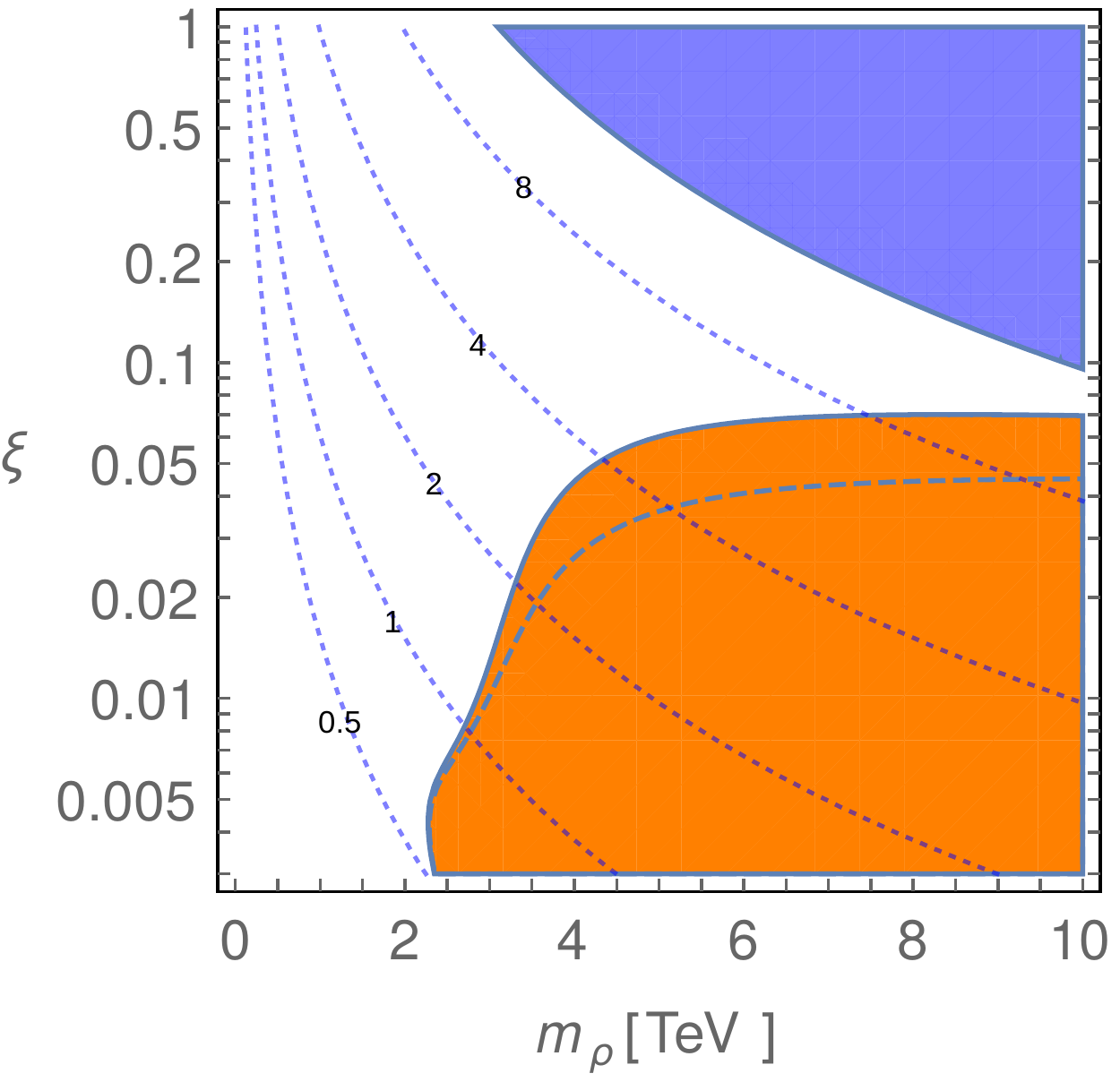}
\hspace{0.1cm}
\end{center}
\vspace{-1.3cm}
\caption{\small
Limits  in the plane $(m_\rho(m_\rho),\xi)$ from a fit to the $\eps_i$. 
The parameter $\xi$ controls the degree of vacuum misalignment and is related to the decay constant $f$ 
as in Eq.~(\ref{eq:xidef}): $\xi \equiv \sin^2\theta = (v/f)^2$.
On the left: Scenario 1 with $a_\rho(m_\rho) = 1/\sqrt{2}$; 
On the right: Scenario 2 with $a_\rho(m_\rho) = 1$. Both plots are done fixing $\Lambda = 3 m_\rho(m_\rho)$.
The orange area denotes the region allowed at $95\%$ CL from the  1-loop results of Eqs.~(\ref{eq:deps1final})-(\ref{eq:deps3final}). 
The dashed line shows the corresponding limit obtained by including the effect of the $\rho$ at the tree level. The dotted blue lines are 
isocurves of constant $g_\rho(m_\rho)$, and the blue region corresponds to $g_\rho(m_\rho) > 4\pi$.
}
\label{fig:limits}
\end{figure}

Our results are expressed as $95\%$ CL exclusion regions in the plane $(m_\rho(m_\rho),\xi)$.
The left and right plots in Figure~\ref{fig:limits} show the limits respectively for Scenario 1 with $a_\rho(m_\rho) = 1/\sqrt{2}$ (two-site limit) and Scenario 2 with 
$a_\rho(m_\rho) = 1$. 
%
%
Notice that  the tree-level shift to $\eps_3$ is the same in the two cases: $\Delta\eps_3|_\text{tree} = (m_W^2/m_\rho^2)(1-4\alpha_2 g_\rho^2)$ 
(see Eq.~(\ref{eq:deps3final})). In both cases we fix $\Lambda = 3 m_\rho(m_\rho)$ and 
set $\alpha_2(m_\rho)= a_\rho^2 (1-a_\rho^2)/(96\pi^2) \log(m_\rho/\Lambda)$.
This one-loop value is chosen so that $\alpha_2$ vanishes at the scale $\mu = \Lambda$ in the unitary gauge.
The orange area represents the  region allowed at 95\% CL following from the full 1-loop
results of Eqs.~(\ref{eq:deps1final})-(\ref{eq:deps3final}). The dashed line shows instead the corresponding limit obtained by including the effect
of the $\rho$ at the tree level.
The dotted  blue lines are isocurves of constant $g_\rho(m_\rho)$, and the blue area corresponds to the region with $g_\rho(m_\rho) > 4\pi$. 
As expected, the 1-loop $\rho$ contribution is more important for larger values of $g_\rho$, for which the tree-level shift to $\eps_3$ is smaller. 
It gives a negative shift to $\eps_3$ and a small correction
to $\eps_1$, thus enlarging the allowed region.  
The numerical values are reported in Table~\ref{tab:depsi} and compared to the shifts from the IR running and 
 Higgs compositeness. 
%
\begin{table}[tbp]
\centering 
\begin{tabular}{r|cc@{\hspace{2em}}c@{\hspace{2em}}c}
& \multicolumn{2}{c@{\hspace{2em}}}{1-loop $\rho$} & IR & Higgs \\[-0.15cm]
& Scenario 1 & Scenario 2 &  running &  comp. \\
\hline \\[-0.45cm]
$\displaystyle 10^3 \, (0.1/\xi)\times\Delta \eps_1$ & $+0.0041$ & $+0.035$ & $-0.43$ & $+0.057$  \\[-0.1cm]
& $[-0.057, +0.097]$ & $[-0.091, +0.25]$ &  & \\[0.35cm]
$\displaystyle 10^3 \, (0.1/\xi)\times\Delta \eps_3$ & $-0.21$ & $-0.16$ & $+0.16$ & $+0.032$ \\[-0.1cm]
& $[-0.67, -0.14]$ & $[-0.31, -0.032]$ & & 
\end{tabular}
\vspace{0.3cm}
\caption{\small
Corrections to $\eps_1$ and $\eps_3$ in units $10^3 (0.1/\xi)$ from different 1-loop effects: 1-loop $\rho$ contribution 
in Scenario 1 with $a_\rho = 1/\sqrt{2}$ and Scenario 2 with $a_\rho = 1$ obtained by fixing $\Lambda/m_\rho =3$ and neglecting the effect of $\alpha_{1,2}$; 
IR running from $m_\rho = 3\,$TeV to $m_Z$; long-distance contribution from 
Higgs compositeness.
The values in squared parentheses indicate the range of the 1-loop $\rho$ contribution obtained by varying $0.5 < a_\rho < 1.5$ in Scenario 1 and 2.
}
\label{tab:depsi}
\end{table}
%
The effect of including the new physics correction to~$\eps_2$ is small, except for $g_\rho \lesssim 1.5$ where it makes the bound on $m_\rho$ less strong
(tail of the orange region at smaller values of $m_\rho$ and $\xi$).
For small $g_\rho$ the 1-loop $\rho$ contribution  becomes less important and the  limit almost coincides  
with the tree-level one.
The interpretation of our results for very  large values of $g_\rho$ requires some caution:
naively the perturbative expansion breaks down
for $g_\rho \gtrsim 4\pi$ (blue region), but in practice higher-loop effects can become sizable earlier, invalidating our approximate result.
For example, we find that the 1-loop correction to $g_\rho$ and to the pole mass $m^\text{pole}_\rho$ 
becomes as large as the tree-level term already for $g_\rho \sim 5-6$.~\footnote{It is because of the premature loss of perturbativity in the pole mass
that we prefer to show the plots of Fig.~\ref{fig:limits} in terms of the running mass $m_\rho$ rather than in terms of $m^\text{pole}_\rho$.}
Also notice that, as a consequence of fixing $\Lambda/m_\rho =3$,
values $g_\rho > 4\pi/(3 a_\rho)$ correspond to a cutoff scale $\Lambda$ larger than its naive upper limit $4\pi f$. 
The latter should not be interpreted as a sharp bound but rather as an indicative values suggested by NDA.
Yet, the above estimate also suggests that perturbativity might be lost for $g_\rho$ somewhat smaller than $4\pi$.

\begin{figure}[tbh]
\begin{center}
\includegraphics[width=0.48\textwidth]{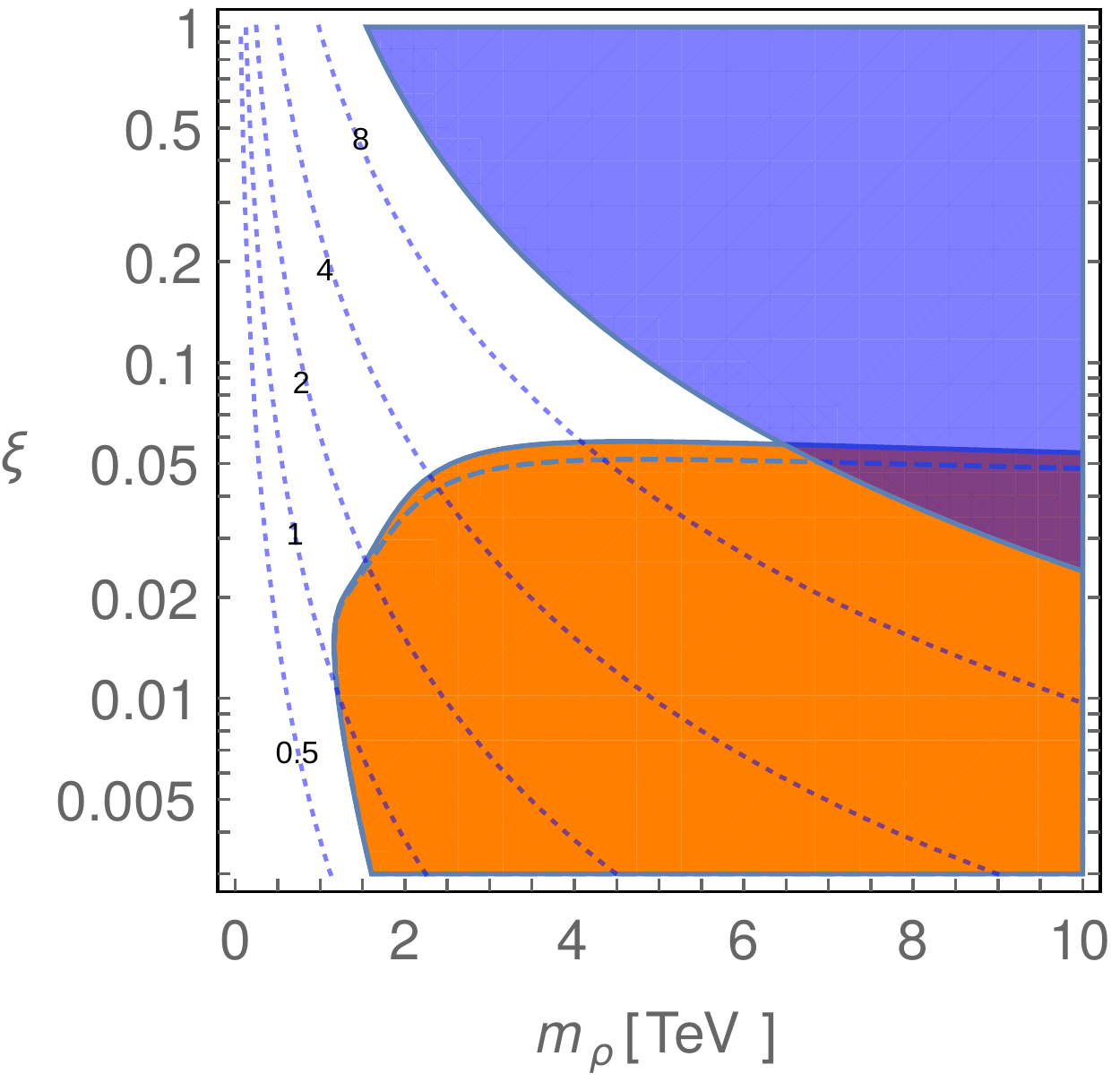}
\hspace{0.4cm}
\includegraphics[width=0.48\textwidth]{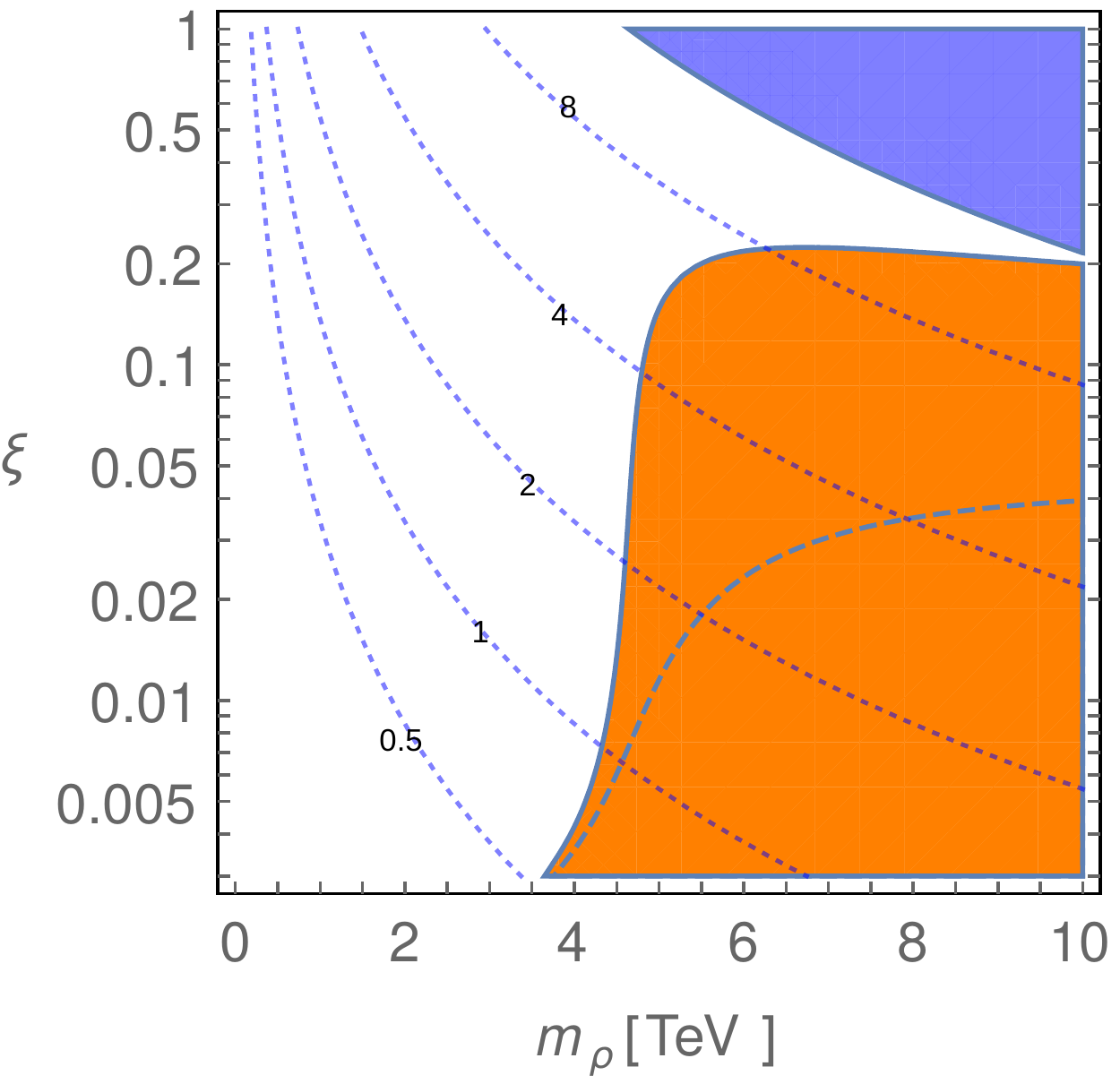}
\hspace{0.1cm}
\end{center}
\vspace{-1.5cm}
\caption{\small
Limits in the plane $(m_\rho(m_\rho),\xi)$ for Scenario 2 with $a_\rho = 0.5$ (left plot) and $a_\rho = 1.5$ (right plot). 
The parameter $\xi$ controls the degree of vacuum misalignment and is related to the decay constant $f$ 
as in Eq.~(\ref{eq:xidef}): $\xi \equiv \sin^2\theta = (v/f)^2$.
Both plots are done fixing $\Lambda = 3 m_\rho(m_\rho)$.
The interpretation of the various curves and regions is the same as in Fig.~\ref{fig:limits}.
}
\label{fig:limitsarho}
\end{figure}

\begin{figure}[tbh]
\begin{center}
\vspace*{0.2cm}
\includegraphics[width=0.48\textwidth]{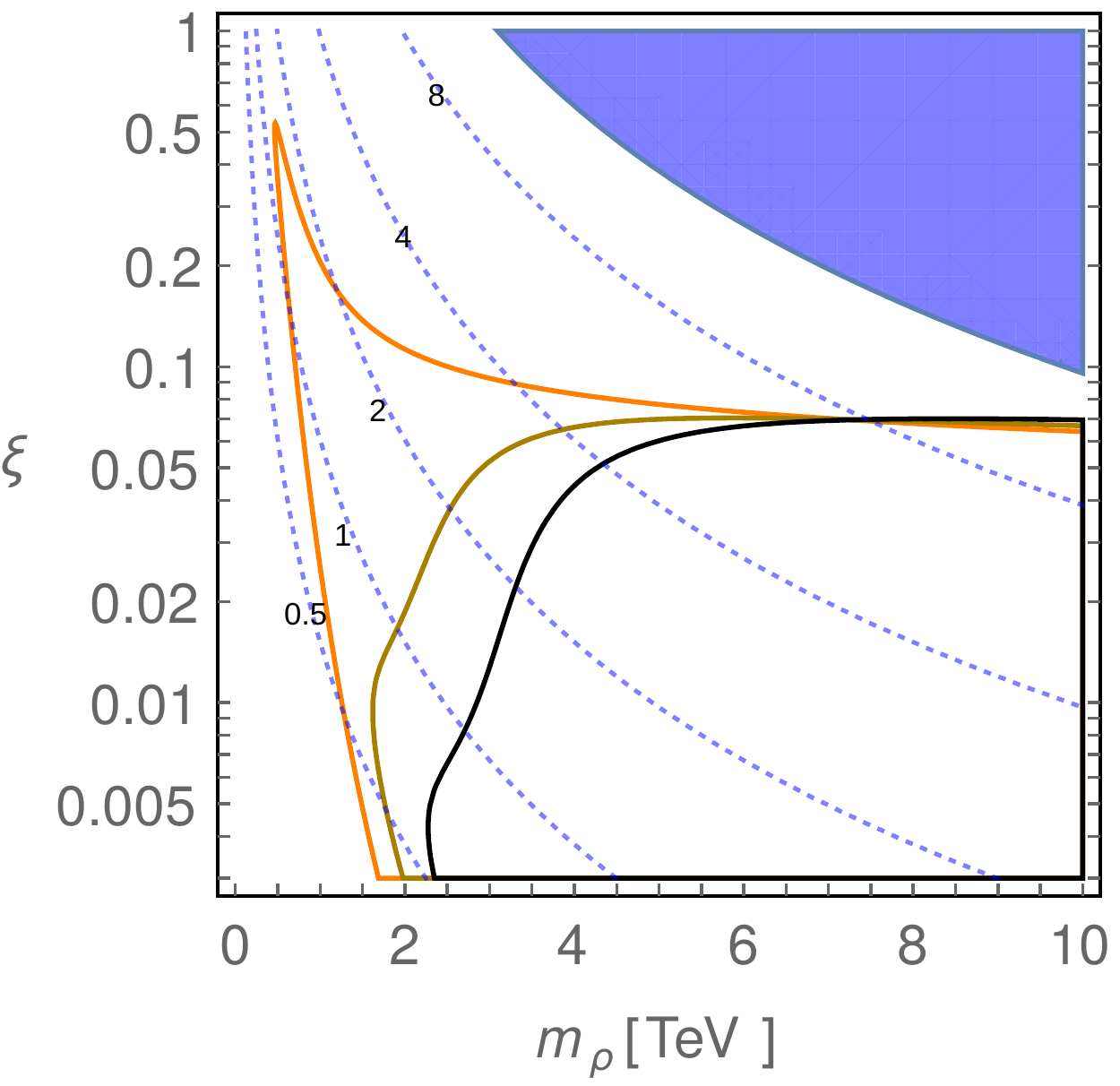}
\end{center}
\vspace{-0.6cm}
\caption{\small
Limits in the plane $(m_\rho(m_\rho),\xi)$ for Scenario 2 with $a_\rho = 1$ and $\Lambda = 3 m_\rho(m_\rho)$.
The parameter $\xi$ controls the degree of vacuum misalignment and is related to the decay constant $f$ 
as in Eq.~(\ref{eq:xidef}): $\xi \equiv \sin^2\theta = (v/f)^2$.
The brown and orange curves are obtained by fixing respectively $\alpha_2 g_\rho^2 = 1/8$ and $1/4$ at the scale $\mu = m_\rho$;
the black curve refers to the case $\alpha_2(\Lambda) =0$ and corresponds to the limit shown in the right plot of Fig.~\ref{fig:limits}.
The region below each curve is allowed at $95\%$ CL.
The dotted blue lines are isocurves of constant $g_\rho(m_\rho)$, and the blue region corresponds to $g_\rho(m_\rho) > 4\pi$.
}
\label{fig:limitsalpha2}
\end{figure}

The plots of Figure~\ref{fig:limits} shows the limits for a benchmark choice of parameters.  When these latter are varied, the 
results can change even significantly.  Increasing the value of the gap $\Lambda/m_\rho$ amplifies the logarithmic term in the 1-loop $\rho$ contribution.
For values of the other parameters as in Fig.~\ref{fig:limits}, the effect turns out to be small and tends to reduce the allowed region. 
Varying $a_\rho$ has a larger impact on the fit, since this parameter controls the size of the tree-level correction to $\eps_3$: smaller values
of $a_\rho$ imply smaller $\Delta\eps_3|_\text{tree}$, hence weaker bounds on $m_\rho$. The value of $a_\rho$ also controls the size and the sign of the
1-loop $\rho$ contribution. Table~\ref{tab:depsi} shows for example how this changes when varying $0.5 < a_\rho < 1.5$.
We find that in general the finite part is numerically comparable, if not larger, than the log term. 
For illustration we show in Figure~\ref{fig:limitsarho} the limits obtained
in Scenario 2 for $a_\rho = 0.5$ (left plot) and $a_\rho = 1.5$ (right plot). 
%
%
Finally, one could consider a scenario where $\alpha_2$  is of order $1/g_\rho^2$,  leading to a cancellation in the tree-level contribution to 
$\eps_3$.~\footnote{A scenario of this kind, with $\alpha_1 \ll \alpha_2 \sim 1/g_\rho^2$, does not satisfy the PUVC criterion,
since the latter requires $\alpha_1- \alpha_2 \lesssim 1/(g_* g_\rho)$.}
A proper calculation of the $\Delta\eps_i$ in this case requires including the 1-loop contribution from $\alpha_2$ through the formulas of Appendix~\ref{sec:alpha2oneloop},
thus re-summing all powers of $\alpha_2 g_\rho^2$.
As an illustration, Figure~\ref{fig:limitsalpha2} shows the limits obtained for $\alpha_2 g_\rho^2 = 1/8$ and $1/4$ at the scale $\mu = m_\rho$,
corresponding respectively to a $50\%$ and $100\%$ cancellation of the tree-level contribution to $\eps_3$.
%
%
In the (extreme) case of a complete cancellation, the tail of the allowed region at large $\xi$ and small $m_\rho$ is a result of the new physics 
contribution to $\eps_2$. It is indeed possible to compensate the positive (negative) shift to $\eps_3$ ($\eps_1$) from the IR running
with a sizable and negative $\Delta\eps_2$, due to the correlation in the 3-dimensional $\chi^2$ function. For small $g_\rho$ such large and negative 
$\Delta\eps_2$ is provided by the tree-level $\rho$ exchange, thus leading to the narrow region extending up to $\xi \sim 0.5$ and $m_\rho \sim 500\,$GeV.

\begin{figure}[tp]
\begin{center}
\includegraphics[width=0.45\textwidth]{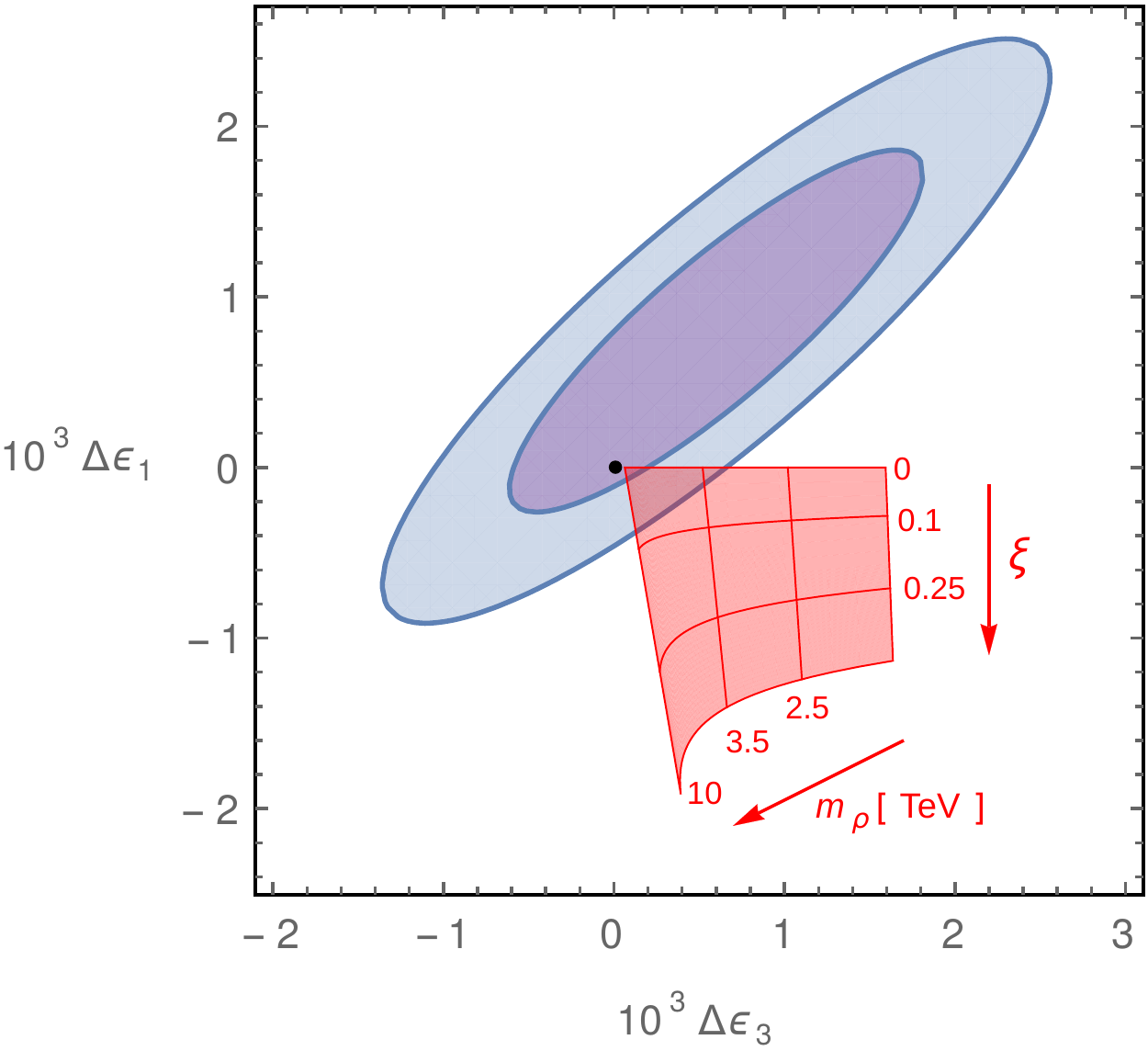} \\[0.5cm]
\includegraphics[width=0.45\textwidth]{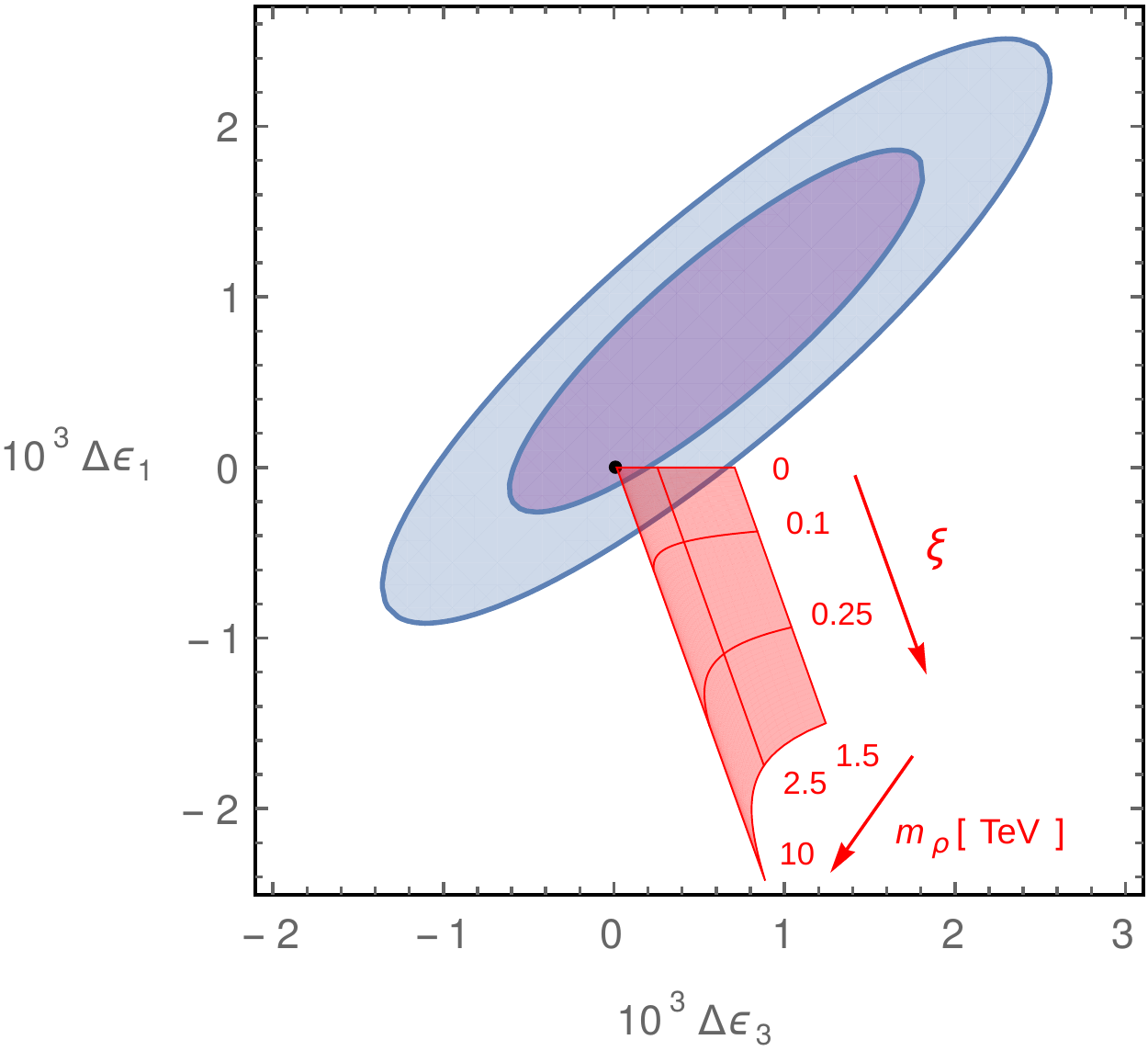}
\includegraphics[width=0.45\textwidth]{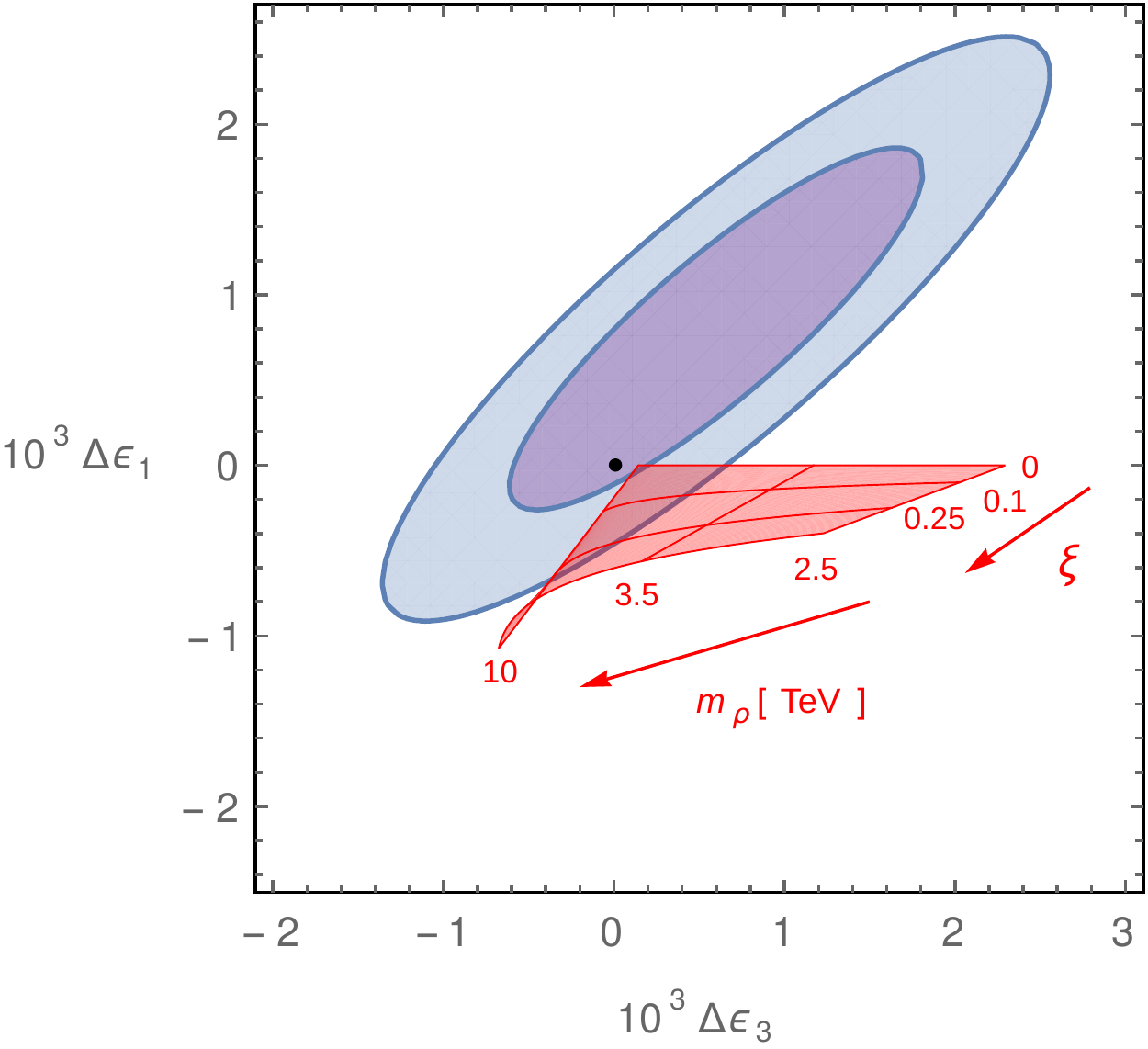}
\hspace{0.1cm}
\end{center}
\vspace{-0.5cm}
\caption{\small
Comparison between the experimental determination of $\Delta\eps_1$, $\Delta\eps_3$ (blue ellipses at $68\%$ and $95\%$ CL) and the theoretical prediction 
in our model (red area). This latter is obtained for the case of Scenario 2 by fixing $a_\rho$ and varying $\xi$ and $m_\rho$ as follows:
$a_\rho = 1$, $\xi = 0-0.4$, $m_\rho = 2-10\,$TeV  (upper plot); $a_\rho = 0.5$, $\xi = 0-0.4$, $m_\rho = 1.5-10\,$TeV (lower left plot);
$a_\rho = 1.5$, $\xi = 0-0.4$, $m_\rho = 2.5-10\,$TeV (lower right plot). The black dot indicates the SM point.
All plots have been obtained by fixing $\eps_2$ to its SM value.
}
\label{fig:eps13}
\end{figure}

The bounds that follow on $m_\rho$ and $\xi$ from our analysis are quite severe.  
As already pointed out in previous studies, this is  because the tree-level exchange of the $\rho$  generally implies a large and positive $\Delta\eps_3$,
while the IR running gives a positive $\Delta\eps_3$ and a negative $\Delta\eps_1$. The combination of these effects brings the theoretical prediction
far outside the $95\%$~CL contour in the plane $(\eps_3,\eps_1)$ unless $\xi$ ($m_\rho$) is very small (large).  This is illustrated by  Figure~\ref{fig:eps13}, where the
region spanned by varying $m_\rho$ and $\xi$ is shown in red for $a_\rho = 0.5,1,1.5$ in the case of Scenario~2. 
%
%
It is evident that an additional negative contribution to $\eps_3$ or positive contribution to $\eps_1$, as for example coming from
loops of fermionic resonances, can relax even significantly the bounds (see for example Refs.~\cite{Grojean:2013qca,Azatov:2013ura})

\section{Conclusions}
\label{sec:conclusions}

In this paper we have computed the 1-loop contribution to the electroweak parameters $\eps_{1,2,3}$ arising from spin-1 resonances
in a class of $SO(5)/SO(4)$ composite Higgs theories. We performed our analysis by giving a low-energy effective description of the strong dynamics
in terms of Nambu-Goldstone bosons and lowest-lying spin-1 resonances ($\rho^L$ and $\rho^R$), 
these latter transforming as an adjoint representation of the unbroken $SO(4)$.
We provided a classification of the relevant operators by including the custodially-breaking effects arising from the external gauging of hypercharge.
A detailed discussion was given of the so-called `hidden local symmetry' description of the spin-1 resonances, where their longitudinal polarizations are parametrized
in terms of the NG bosons from a larger coset. This was useful to analyze a particular limit, noticed by Ref.~\cite{Panico:2011pw}, in which the theory acquires a larger 
$SO(5)\times SO(5)/SO(5)$ global symmetry and has a collective breaking mechanism. In particular, we reviewed the argument that shows how certain EWSB quantities
enjoy an improved convergence in this limit, clarifying the role of divergent subdiagrams in the calculation of $S$ and $T$.

The contribution of the $\rho$ to the electroweak parameters was computed by performing a 1-loop matching to the low-energy theory 
of NG bosons. We used dimensional regularization and analyzed in detail the renormalization of the spin-1 Lagrangian and the RG evolution 
of its coefficients. 
We estimate a relative uncertainty in our calculation  of order $m_h^2/m_\rho^2$ from neglecting 
the EW and Higgs boson masses 
in the matching and  truncating the effective Lagrangian at leading order in the derivative expansion,
and of order $g^2/g_\rho^2$ from neglecting diagrams with additional insertions of the elementary vector bosons.
Our results extend previous studies where the contribution from spin-1 resonances was included only at the tree level.
They represent a starting point for a complete 1-loop analysis including all the lowest-lying resonances, in particular the top partners.

By including only the spin-1 resonances, a fit to the current electroweak data gives rather strong bounds. 
We find that typical $95\%$ probability limits on the $\rho$ mass and  the degree of Higgs compositeness are in the range $m_\rho \gtrsim 3-4\,$TeV 
and $\xi \lesssim 0.1-0.05$,  although choices of parameters exist which lead to less stringent constraints.
The 1-loop  contribution from the $\rho$ can be most easily evaluated by expressing the $\Delta\eps_i$ in terms of the parameters of the spin-1 Lagrangian renormalized
at the scale $m_\rho$ (Eqs.~(\ref{eq:deps1final}-\ref{eq:deps3final})). Although parametrically subdominant compared to the IR running and the tree-level contribution, 
we find it to be numerically  important in a significant fraction of the parameter space, where the coupling strength $g_\rho$ is moderately large.
Its effect is that of enlarging the allowed region  providing a negative shift to $\eps_3$ (see Table~\ref{tab:depsi} and Figs.~\ref{fig:limits}-\ref{fig:limitsalpha2}).
The relative importance of the 1-loop contribution grows with $g_\rho$.
Although one would naively expect perturbativity to remain valid until $g_\rho \sim 4\pi$,  the 1-loop correction becomes as important as the tree-level 
term already for $g_\rho \sim 5-6$ in several quantities, as for example
the running of $g_\rho$ or the pole mass $m_\rho^\text{pole}$.
This suggests that any limit extending to such large values of  $g_\rho$ should be interpreted with caution.
The  contribution from cutoff states to the electroweak observables might also be important. Its naive estimate in the case of a fully strongly coupled
dynamics at the scale $\Lambda$ suggests that it is subleading compared to the 1-loop $\rho$ contribution only by a factor $\log(\Lambda/m_\rho)$, 
which is not expected to be very large.
In fact, the very existence of a gap $\Lambda/m_\rho \gg 1$ should be considered as a working hypothesis of our study, 
not necessarily realized by the underlying strong dynamics. In this sense our calculation should be regarded
as a way, more refined than a simple estimate, to assess the contribution  of the spectrum of resonances 
lying at the compositeness scale to the  oblique parameters.

\section*{Acknowledgments}

We would like to thank 
Marco Bochicchio,
Enrico Franco,
Davide Greco,
Gino Isidori,
Giuliano Panico,
Riccardo Rattazzi,
Slava Rychkov,
Luca Silvestrini,
Massimo Testa,
Enrico Trincherini
and Andrea Wulzer
for useful discussions.
The work of R.C. was partly supported by the ERC Advanced Grant No.~267985 
\textit{Electroweak Symmetry Breaking, Flavour and Dark Matter: One Solution for Three Mysteries (DaMeSyFla)}.

\appendix
\section*{Appendix}
\section{Two-site Lagrangian in the $SO(5)\times SO(5)_H$ limit}
\label{app:two-site}

As discussed in Section~\ref{sec:effeLagrangian}, in the limit $a_\rhoL = a_\rhoR = 1/\sqrt{2}$ the Lagrangian~(\ref{eq:kinterms}) enjoys a larger 
$SO(5) \times SO(5)_H \to SO(5)_d$ global symmetry, partially gauged by the EW and $\rho_\mu$ fields.
The theory is in fact equivalent to a two-site $SO(5) \times SO(5)_H$ model where $W_\mu$ and $B_\mu$ gauge a subgroup $SU(2)_L \times U(1)_Y$
on the left site, while $\rho_\mu$ gauges an $SO(4)_H$ on the right site.
The most convenient way to construct the Lagrangian, in this case, is in terms of a $5 \times 5$ link field 
$\bar U(\pi, \eta) = e^{ i \sqrt{2}\, \pi(x)/f } e^{i \sqrt{2}\, \eta(x)/f}$,
where $\pi(x) = \pi^{\hat a}(x) T^{\hat a} $, $\eta(x) =  \eta^a(x) T^a$ and  $T^{\hat a}$, $T^a$ are the $SO(5)$ generators.
The link transforms as a $(5, \bar 5)$ under $SO(5) \times SO(5)_H$ 
\begin{equation}
\bar U(\pi,\eta) \to g \, U(\pi, \eta) \, g^\dagger_H \, ,
\end{equation}
so that its covariant derivative is (we conveniently normalize gauge fields so that  gauge couplings appear in their kinetic terms)
\begin{equation}
D_\mu \bar U = \partial_\mu \bar U + i W_\mu^{a_L} T^{a_L} \bar U + i B_\mu T^{3_R} \bar U - i \bar U \rho_\mu^a T^a\, .
\end{equation}
Given the above transformation rules, it is possible to eat all the NG bosons $\eta$ by making an $SO(4)_H$ local transformation and go to a gauge
in which the link field coincides with $U(\pi)$ defined in Section~\ref{sec:effeLagrangian}:  $\bar U(\pi,\eta=0) = e^{ i \sqrt{2}\, \pi(x)/f } = U(\pi)$.
When acting on $\bar U$ from the left with a global rotation $g \in SO(5)$, the unitary gauge can be maintained by simultaneously performing  a
suitable, local $SO(4)_H$ transformation $g_H = h(g , \pi)$. The fields thus obey the following transformation rules
\begin{equation}
\begin{split}
\bar U(\pi,0) & \to \bar U(g(\pi), 0) = g \, \bar U(g,0) h^\dagger(g,\pi) \\[0.2cm]
\rho_\mu & \to h(g,\pi) \rho_\mu h^\dagger(g,\pi) - i h(g,\pi) \partial_\mu h^\dagger(g,\pi) \, ,
\end{split}
\end{equation}
which are the same as those in the $SO(5)/SO(4)$ theory with massive spin-1 resonance $\rho_\mu$.

By working in the $\eta=0$ gauge, it is easy to recast the kinetic term of $\bar U$ in $SO(5)/SO(4)$ CCWZ notation.
Since $-i \bar U(\pi, 0) D_\mu \bar U(\pi,0) = d_\mu(\pi) + E_\mu(\pi) -\rho_\mu$, it simply follows
\begin{equation}
\frac{f^2}{4} \Tr\!\left[ (D_\mu \bar U)^\dagger (D_\mu \bar U) \right] = \frac{f^2}{4} \Tr\!\left[ d_\mu(\pi) d^\mu(\pi) \right] + 
\frac{f^2}{4} \Tr\!\left[ (\rho_\mu - E_\mu(\pi))^2 \right] \, ,
\end{equation}
which gives $a_\rho = 1/\sqrt{2}$ upon comparison with  Eq.~(\ref{eq:Lrho}).

At the level of two derivatives and two powers of the hypercharge spurion $g' T^{3_R}_0$, there is one $(SO(5)\times SO(5)_H)$-invariant operator which can be
constructed:
\begin{equation}
\bar O_T = \left( \Tr\!\left[ \bar U i D_\mu \bar U^\dagger g' T^{3_R}_0 \right] \right)^2 \, .
\end{equation}
Notice that the combination $\bar U D_\mu \bar U^\dagger$ transforms as $\bar U D_\mu \bar U^\dagger \to g (\bar U D_\mu \bar U^\dagger) g^\dagger$.
In the $\eta = 0$ gauge, by defining $\chi(\pi) = \bar U^\dagger(\pi,0) g' T^{3_R}_0 \bar U(\pi,0)$, one has
\begin{equation}
\bar O_T = \left( \Tr\!\left[ (d_\mu + E_\mu(\pi) - \rho_\mu ) \chi \right] \right)^2
\end{equation}
which coincides with the right-hand side of Eq.~(\ref{eq:OTsymm}).
On the other hand, at order $g_\rho^0$ there is no  operator with two EW field strengths and no derivative acting on $\bar U$ which can contribute to the $S$ parameter. 
This is because there is no way to saturate the $SO(5)_H$ index of $\bar U$ except in the trivial product $\bar U\bar U^\dagger = 1$.

\section{Functions $f_{1,2,3}$}
\label{app:formulas}

We report here the expressions of the functions $f_{1,2,3}$ of Eqs.~(\ref{eq:eps1})-(\ref{eq:eps3}) that parametrize the
1-loop Higgs contribution to the $\eps_i$:
\begin{align}
\begin{split}
f_{1}(h) & =\frac{1}{s_W^{2}}\left(-\frac{5c_W^{2}}{12}+\frac{h^{2}}{6}-\frac{7h}{12}+\frac{31}{18}\right) \\
           & - \frac{\log(h)}{12s_W^{2}\left(c_W^{2}-h\right)} \left[\left(c_W^{2}+5\right)h^{3}-\left(5c_W^{2}+12\right)h^{2}+2\left(9c_W^{2}+2\right)h-4c_W^{2}-h^{4}\right]\\
           & -\frac{c_W^{4}}{s_W^{2}\left(h-c_W^{2}\right)}\log(c_W)+\frac{h\left(h^{3}-7h^{2}+20h-28\right)}{6s_W^{2}\sqrt{(4-h)h}}\, \arctan\left(\sqrt{\frac{4}{h}-1}\right) \, ,
\end{split} \\[0.7cm]
\begin{split}
f_{2}\left(h\right) & =\left(-\frac{1}{c_W^{4}}-2\right)h^{2}+\left(\frac{9}{2c_W^{2}}+6\right)h-\frac{47}{2} \\
                           & +\frac{\log(c_W)}{c_W^{6}\left(c_W^{2}-h\right)} \left(2c_W^{8}-38c_W^{6}h+24c_W^{4}h^{2}-7c_W^{2}h^{3}+h^{4}\right) \\
                           & + \frac{\log(h)}{2c_W^{6}\left(c^{2}-h\right)}  \Big[-12c_W^{8}-\left(2c_W^{6}+1\right)h^{4}+6\left(3c_W^{2}+8\right) c_W^{6}h \\
                           &  \hspace{3.0cm}-3\left(3c_W^{4}+6c_W^{2}+8\right)c_W^{4}h^{2}+\left(2c_W^{6}+9c_W^{4}+7\right)c_W^{2}h^{3}\Big] \\
                           & -\frac{\left(2h^{3}-13h^{2}+32h-36\right)h}{\sqrt{(4-h)h}}\, \arctan\left(\sqrt{\frac{4}{h}-1}\right) \\
                           & +\frac{\left(48c_W^{6}h-28c_W^{4}h^{2}+8c_W^{2}h^{3}-h^{4}\right)}{c_W^{6}\sqrt{h\left(4c_W^{2}-h\right)}}\, \arctan\left(\sqrt{\frac{4c_W^{2}}{h}-1}\right) \, , 
\end{split}\\[0.7cm]
\begin{split}
f_{3}\left(h\right) & =\left(-h^{2}+3h-\frac{31}{6}\right)+\frac{1}{4}\left(2h^{3}-9h^{2}+18h-12\right)\log(h) \\
                           & -\frac{\left(2h^{3}-13h^{2}+32h-36\right)h}{2\sqrt{(4-h)h}}\, \arctan\left(\sqrt{\frac{4}{h}-1}\right)\, .
\end{split}
\end{align}
They agree with the functions $H_i$ of Ref.~\cite{Orgogozo:2012ct}, see also Ref.~\cite{Novikov:1992rj}.

\section{One-loop renormalization of the spin-1 Lagrangian}
\label{sec:spin1renorm}

Consistently with the 1-loop matching of the full and effective theories, one should also perform a 1-loop renormalization of the 
Lagrangian of spin-1 resonances.  We first describe our procedure for the unitary gauge and then give the results also for the Landau gauge.
We will not specify the quantum numbers of the spin-1 resonance unless necessary since the same expressions hold for both $\rho_L$ and $\rho_R$,
there being no mixed renormalization at one loop.

Starting from the bare Lagrangian, we define renormalized fields and parameters as follows
\begin{equation}
\begin{split}
\pi^{\hat a(0)} & = Z_\pi^{1/2} \pi^{\hat a}  \\
\rho^{a (0)}_\mu & = Z_\rho^{1/2} \rho^{a}_\mu \\
W^{i (0)}_\mu & = Z_W^{1/2} W^{i}_\mu \\
B^{(0)}_\mu & = Z_B^{1/2} B_\mu
\end{split}
\hspace{1.5cm}
\begin{split}
f^{(0)} & = \mu^{-\eps/2}  Z_f^{1/2} f(\mu) \\
m_\rho^{(0)} & = Z_m m_\rho(\mu) \\
g_\rho^{(0)} & = \mu^{\eps/2} Z_{g_\rho} g_\rho(\mu) \\
g^{(0)} & = \mu^{\eps/2} Z_g g(\mu) \\
g^{\prime (0)} & = \mu^{\eps/2} Z_{g'} g'(\mu) \, ,
\end{split}
\end{equation}
where $Z_i$ are renormalization functions and
we make use of dimensional regularization in $d = 4 - \eps$ dimensions with a renormalization scale $\mu$.
The renormalization of the elementary gauge fields and coupling constants arises at $O(g^2, g'^2)$ 
so we can  set $Z_W, Z_B, Z_g$ and  $Z_{g'}$ to unity  when working at leading order in an expansion in powers of the elementary  couplings.
The remaining  functions $Z_\pi$, $Z_\rho$, $Z_m$, $Z_f$ and $Z_{g_\rho}$ can be computed by renormalizing the 2-point functions
$\langle \pi \pi \rangle$, $\langle \rho_\mu \rho_\nu\rangle$, $\langle A_\mu A_\nu\rangle$ and $\langle \rho_\mu A_\nu \rangle$, where $A_\mu = W_\mu, B_\mu$.
We adopt a subtraction scheme where the above Green functions (and their derivatives) are evaluated at $q^2 =m_\rho^2$ and made finite by removing
their poles in $1/\bar \eps$, where $2/\bar \eps \equiv 2/\eps - \gamma - \log(4\pi)$. This hybrid $\overline{MS}$ on-shell scheme is convenient,
as it requires the same number of counterterms as in the Landau gauge. Performing instead a minimal subtraction on off-shell Green functions would
require further counterterms to remove the  $q^4$ and $q^6$ divergent terms in the $\rho$ propagator.
We thus obtain
\begin{equation}
\begin{gathered}
Z_\rho = 1 -g_{\rho}^{2}\frac{2a_{\rho}^{4}-53}{96\pi^{2}}\, \frac{1}{\bar\epsilon}\, , \qquad 
Z_{g_{\rho}}  = 1+g_{\rho}^{2}\frac{2a_{\rho}^{4}-85}{192\pi^{2}}\, \frac{1}{\bar\epsilon} \, , \qquad 
Z_{m}  = 1+g_{\rho}^{2}\frac{2a_{\rho}^{4}-69}{192\pi^{2}}\, \frac{1}{\bar\epsilon} \\[0.5cm]
Z_{\pi}  = 1 + \left( g_{\rho_L}^{2}\frac{3a_{\rho_L}^{4}}{16\pi^{2}} +g_{\rho_R}^{2}\frac{3a_{\rho_R}^{4}}{16\pi^{2}} \right) \frac{1}{\bar\epsilon}\, , \qquad 
Z_{f}  = 1+ \left( g_{\rho_L}^{2}\frac{9a_{\rho_L}^{4}}{32\pi^{2}} + g_{\rho_R}^{2}\frac{9a_{\rho_R}^{4}}{32\pi^{2}}\right) \frac{1}{\bar\epsilon} \, .
\end{gathered}
\end{equation}
From these expressions it follows Eq.~(\ref{eq:RGgrho}) and
\begin{equation}
\mu \frac{\partial m_\rho}{\partial\mu}  \equiv \beta_{m_\rho} =  g_\rho^2 \frac{2 a_\rho^4 - 69}{192\pi^2} m_\rho \, , \qquad
\mu \frac{\partial f}{\partial \mu}  \equiv \beta_f = f \left( g_{\rho_L}^2 \frac{9 a_{\rho_L}^4}{64\pi^2} + g_{\rho_R}^2 \frac{9 a_{\rho_R}^4}{64\pi^2} \right) \, .
\end{equation}
The renormalized $c_i$ and $\alpha_2$ are instead defined by
\begin{equation}
\begin{split}
c_i^{(0)}   & = \mu^{-\eps} \left( c_i(\mu)  + \frac{1}{\bar\eps} \Delta_{i} \right)\simeq c_i(\mu) + \Delta_i \left( \frac{1}{\bar\eps}  - \log\mu \right) \\[0.3cm]
\alpha_2^{(0)} &  = \mu^{-\eps} \left( \alpha_2(\mu)  + \frac{1}{\bar\eps} \Delta_{\alpha_2} \right) \simeq \alpha_2(\mu) + 
\Delta_{\alpha_2} \left( \frac{1}{\bar\eps}  - \log\mu \right)\, .
\end{split}
\end{equation}
The value of the counterterm $\Delta_{\alpha_2}$ is obtained by renormalizing the $\langle \rho_\mu A_\mu \rangle$ Green function. We find
$\Delta_{\alpha_2} = a_\rho^2 (1-a_\rho^2)/96\pi^2$, which leads to Eq.~(\ref{eq:RGalpha2}).
The value of the counterterms $\Delta_{c_i}$ is instead found by renormalizing the Green functions in Figs.~\ref{fig:S_tree}-\ref{fig:T_loops} 
after canceling the divergences from subdiagrams. The corresponding RG evolution of the coefficients $c_i$ is given in 
Eqs.~(\ref{eq:RGcS}),~(\ref{eq:RGcT}) and~(\ref{eq:RGc3W}).

A similar procedure also applies in the Landau gauge with a few differences however. First, another field is present, that of the NG bosons $\eta$,
which needs to be renormalized. Second, the $\rho$ mass originates from the $\eta$ kinetic term, and $m_\rho$ is defined in terms of $f_\rho$
according to Eq.~(\ref{eq:arho}). It is thus more convenient to include $f_\rho$ in the list of renormalized quantities and treat $m_\rho$ as a derived parameter.
By defining
\begin{equation}
\eta^{a (0)} = Z_\eta^{1/2} \eta^a \, , \quad \qquad  f_\rho^{(0)} = \mu^{-\eps/2}  Z_{f_\rho}^{1/2} f_\rho(\mu)
\end{equation}
we find
\begin{equation}
\begin{gathered}
Z_{\rho}  = 1-g_{\rho}^{2}\frac{2a_{\rho}^{4}-51}{96\pi^{2}}\, \frac{1}{\bar\epsilon} \, , \hspace{0.6cm}
Z_{g_{\rho}}  = 1+g_{\rho}^{2}\frac{2a_{\rho}^{4}-87}{192\pi^{2}}\, \frac{1}{\bar\epsilon}\, ,  \hspace{0.6cm}
Z_{f_{\rho}}   = Z_\eta  = 1+g_{\rho}^{2}\frac{3}{16\pi^{2}}\, \frac{1}{\bar\epsilon} \\[0.5cm]
Z_{\pi}   = 1+ \left( g_{\rho_L}^{2}\frac{a_{\rho_L}^{4}}{4\pi^{2}} + g_{\rho_R}^{2}\frac{a_{\rho_R}^{4}}{4\pi^{2}} \right) \frac{1}{\bar\epsilon}\, , \qquad 
Z_{f}  = 1+ \left( g_{\rho_L}^{2}\frac{9a_{\rho_L}^{4}}{32\pi^{2}} + g_{\rho_R}^{2}\frac{9a_{\rho_R}^{4}}{32\pi^{2}} \right) \frac{1}{\bar\epsilon}
\end{gathered}
\end{equation}
and $\Delta_{\alpha_2} = (2 a_\rho^2 (1-a_\rho^2))/192\pi^2$. The corresponding RG equations read
\begin{equation}
\begin{split}
\mu \frac{\partial g_\rho}{\partial\mu}  & = g_\rho^3 \frac{2 a_\rho^4 - 87}{192\pi^2}\, ,  \\[0.3cm]
\mu \frac{\partial \alpha_2}{\partial \mu} & =  \frac{2 a_\rho^2 (1-a_\rho^2)+1}{192\pi^2}\, , 
\end{split}
\hspace{1.5cm}
\begin{split}
\mu \frac{\partial f_\rho}{\partial\mu}  & =  g_\rho^2 \frac{3}{32\pi^2} f_\rho\, , \\[0.3cm]
\mu \frac{\partial f}{\partial \mu}  & = f \left( g_{\rho_L}^2 \frac{9a_{\rho_L}^4}{64\pi^2} + g_{\rho_R}^2 \frac{9a_{\rho_R}^4}{64\pi^2}  \right) \, .
\end{split}
\end{equation}
%

\section{One-loop contribution from $\alpha_2$}
\label{sec:alpha2oneloop}

When including the effect of $\alpha_2$ at the 1-loop level, there arise the following additional contributions to the $\eps_i$:
\begin{align}
\begin{split}
\Delta\epsilon_{1}\big|_{\alpha_2} = &\, -\frac{9g^{\prime2}}{128\pi^{2}} \sin^{2}\!\theta \\[0.1cm]
& \, \times \Bigg\{
 \frac{8}{3} \frac{a_{\rho_L}^{2} m_{\rho_L}^{2}}{m_{\rho_L}^{2}-m_{\rho_R}^2} 
 \bigg[ \, 8\left(1-\alpha_{2L} g_{\rho_L}^2\right)\alpha_{2L}\alpha_{2R} g_{\rho_L}^2 g_{\rho_R}^2 \left( a_{\rho_R}^{2} - \alpha_{2R} g_{\rho_L}^2 a_{\rho_L}^{2} \right) \\
        & \hspace{3.5cm}  -\left(1-\alpha_{2L} g_{\rho_L}^2\right) \alpha_{2L} g_{\rho_L}^2\left(2a_{\rho R}^{2}+\frac{m_{\rho_R}^{2}}{m_{\rho_L}^{2}}-1\right) \\
        & \hspace{3.5cm}  -2 \alpha_{2R} g_{\rho_R}^2 \left( a_{\rho_R}^{2} - \alpha_{2R} g_{\rho_L}^2 a_{\rho_L}^{2}  \right) \bigg] \log\frac{\mu}{m_{\rho_L}} \\
        & \phantom{\, \times \Bigg\{} + \frac{2}{9}a_{\rho L}^{2} \alpha_{2L} g_{\rho_L}^2
           \bigg[ 11-10a_{\rho R}^{2} + 20\alpha_{2R} g_{\rho_R}^2 a_{\rho R}^{2} \\
           & \hspace{3.5cm} + 20\,  \alpha_{2L} g_{\rho_L}^2 \alpha_{2R} g_{\rho_R}^2 a_{\rho R}^{2} \left(1+\frac{m_{\rho_L}^{2}}{m_{\rho_R}^{2}}+\frac{m_{\rho_R}^{2}}{m_{\rho_L}^{2}}\right) \\
           & \hspace{3.5cm} -40\alpha_{2L} g_{\rho_L}^2 \alpha_{2R} g_{\rho_R}^2 a_{\rho R}^{2} \left(1+\frac{m_{\rho_R}^{2}}{m_{\rho_L}^{2}}\right)  \\
           & \hspace{3.5cm} -\alpha_{2L} g_{\rho_L}^2 \left(11-10a_{\rho_R}^{2}\left(1+\frac{m_{\rho_R}^{2}}{m_{\rho_L}^{2}}\right)\right)\bigg] \Bigg\} 
+ \{ L\leftrightarrow R \}\, ,
\end{split}
\\[0.5cm]
\begin{split}
\Delta\eps_{2} \big|_{\alpha_2} = & \, \frac{g^{2}}{96\pi^{2}} \frac{g^2}{g_{\rho_L}^2} \frac{1}{a_{\rho_L}^2} \sin^{2}\!\theta  \cos^4\frac{\theta}{2}\\[0.1cm]
 &\, \times \Bigg\{
 \log\frac{\mu}{m_{\rho_R}} \left[ 116\, \alpha_{2L} \, g_{\rho_L}^2  - \alpha_{2L}^2 \, g_{\rho_L}^4 \left( 74 - 6 a_{\rho_L}^2 \tan^2\frac{\theta}{2} \right) \right]  \\
 & \phantom{\, \times \Bigg\{} +  \alpha_{2L} \, g_{\rho_L}^2 \left( 5 - 6 a_{\rho_L}^2 \tan^2\frac{\theta}{2} \right) 
    + \alpha_{2L}^2 \, g_{\rho_L}^4 \left( 7 + \frac{17}{2}  a_{\rho_L}^2 \tan^2\frac{\theta}{2} \right)
 \Bigg\} \\
 & \phantom{\, \times \Bigg\{} + \left\{ L\leftrightarrow R, \ \theta \to \pi - \theta \right\} 
\end{split}
\\[0.5cm]
\begin{split}
\Delta\eps_{3} \big|_{\alpha_2} = & \, \frac{g^{2}}{96\pi^{2}}\sin^{2}\!\theta 
  \bigg[
  \, \frac{3}{2} \alpha_{2L}\, g_{\rho_L}^2 \left( 9a_{\rho L}^{2}-4 +\alpha_{2L} g_{\rho_L}^2 \left(9a_{\rho L}^{2}-8\right)\right) \\
  & \phantom{\, \frac{g^{2}}{96\pi^{2}}\sin^{2}\!\theta \bigg\{ \, }
   + 18\left(\alpha_{2L} g_{\rho_L}^2 \left(a_{\rho L}^{2}+2\right)-\alpha_{2L}^{2}a_{\rho L}^{4}\right) \log\frac{\mu}{m_{\rho_L}}
 \bigg]+ \{ L\leftrightarrow R \}\, .
\end{split}
\end{align}
The renormalization of the various parameters is also affected, in particular each $\beta$-function gets an additional contribution.
We report the corresponding expressions in the unitary gauge:
\begin{align}
\begin{split}
\Delta \beta_{c_3^+} = &  -\alpha_{2L}g_{\rho_L}^{2} \frac{2a_{\rho_L}^{4}-20a_{\rho_L}^{2}+11}{192\pi^{2}}+\alpha_{2L}^{2}g_{\rho_L}^{4}\frac{3a_{\rho_L}^{4}-7a_{\rho_L}^{2}+6}{96\pi^{2}}
                                       \\ & -\alpha_{2L}^{3}g_{\rho_L}^{6}\frac{a_{\rho_L}^{4}}{12\pi^{2}} + \{ L \leftrightarrow R \}
\end{split} \\[0.3cm]
\begin{split}
\Delta \beta_{c_T} = &  -\frac{3}{32\pi^{2}}  \frac{a_{\rho_L}^{2} m_{\rho_L}^{2}}{m_{\rho_L}^{2}-m_{\rho_R}^2}  \\
        & \times \bigg[ \, 8\left(1-\alpha_{2L} g_{\rho_L}^2\right)\alpha_{2L}\alpha_{2R} g_{\rho_L}^2 g_{\rho_R}^2 \left( a_{\rho_R}^{2} - \alpha_{2R} g_{\rho_L}^2 a_{\rho_L}^{2} \right) \\
        & \phantom{\times \bigg[ \,}  -2 \alpha_{2R} g_{\rho_R}^2 \left( a_{\rho_R}^{2} - \alpha_{2R} g_{\rho_L}^2 a_{\rho_L}^{2}  \right) \\
        & \phantom{\times \bigg[ \,}  -\left(1-\alpha_{2L} g_{\rho_L}^2\right) \alpha_{2L} g_{\rho_L}^2\left(2a_{\rho R}^{2}+\frac{m_{\rho_R}^{2}}{m_{\rho_L}^{2}}-1\right) \bigg]+ \{ L \leftrightarrow R \}
\end{split} \\[0.3cm]
\Delta \beta_{c_{2W}} = & -\frac{1}{m_{\rho_L}^{2}}\left(\alpha_{2L}g_{\rho_L}^{2}\frac{2a_{\rho_L}^{2}-85}{48\pi^{2}}
                                      +\alpha_{2L}^{2}g_{\rho_L}^{4}\frac{37-3a_{\rho_L}^{2}\tan^{2}(\theta/2)}{24\pi^{2}}\right) \\[0.3cm]
\Delta \beta_{c_{2B}} = & -\frac{1}{m_{\rho_R}^{2}}\left(\alpha_{2R}g_{\rho_R}^{2}\frac{2a_{\rho_R}^{2}-85}{48\pi^{2}}+\alpha_{2R}^{2}g_{\rho_R}^{4}
                                      \frac{37-3a_{\rho_R}^{2}\cot^{2}(\theta/2)}{24\pi^{2}}\right) \\[0.8cm]
\Delta \beta_{g_\rho}  = & - \frac{\alpha_{2}g_{\rho}^{5}}{24\pi^{2}}  \left( a_{\rho}^{4}-a_{\rho}^{2}-3+ \alpha_{2} g_{\rho}^{2} a_{\rho}^{4} \right) \\[0.3cm]
\Delta \beta_{m_\rho}  = & m_{\rho} \, \alpha_{2} g_{\rho}^{4} \frac{a_{\rho}^{4} }{24\pi^{2}} \left( -1 +\alpha_{2} g_{\rho}^{2} \right) \\[0.3cm]
\Delta \beta_{\alpha_2}= & \alpha_{2}g_{\rho}^{2}\frac{4a_{\rho}^{4}-4a_{\rho}^{2}+25}{96\pi^{2}}\, .
\end{align}  
%

\section{Alternative matching for $\tilde c_T$}
\label{sec:altmatchingcT}

\begin{figure}
\begin{center}
\includegraphics[width=0.28\textwidth]{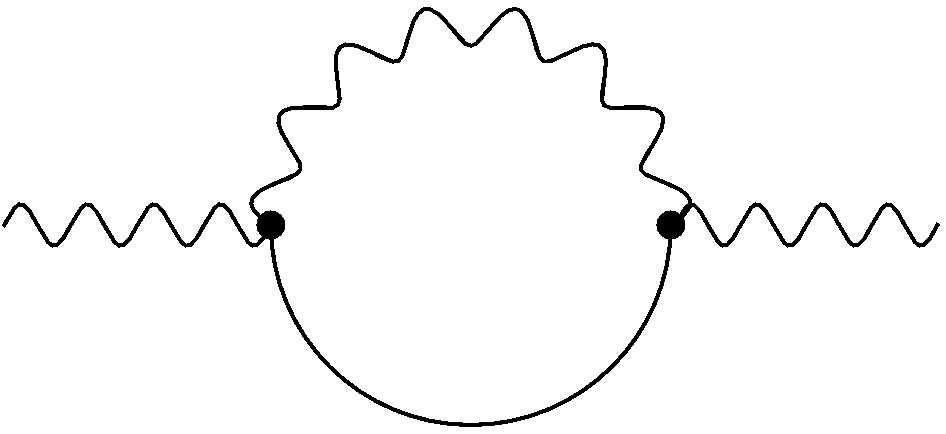}
\end{center}
\caption{
One-loop diagram with NG bosons contributing to the $\langle W^1 W^1 \rangle - \langle W^3 W^3 \rangle$ Green function. 
External (internal) wavy lines denote the elementary $W$ ($B$) field, while continuous lines stand for the NG bosons ($\pi^{\hat a}$ and $\eta$).
}
\label{fig:altT_onlyNGbosons}
\end{figure}

\begin{figure}
\begin{center}
\hspace{0.2cm}
\includegraphics[width=0.28\textwidth]{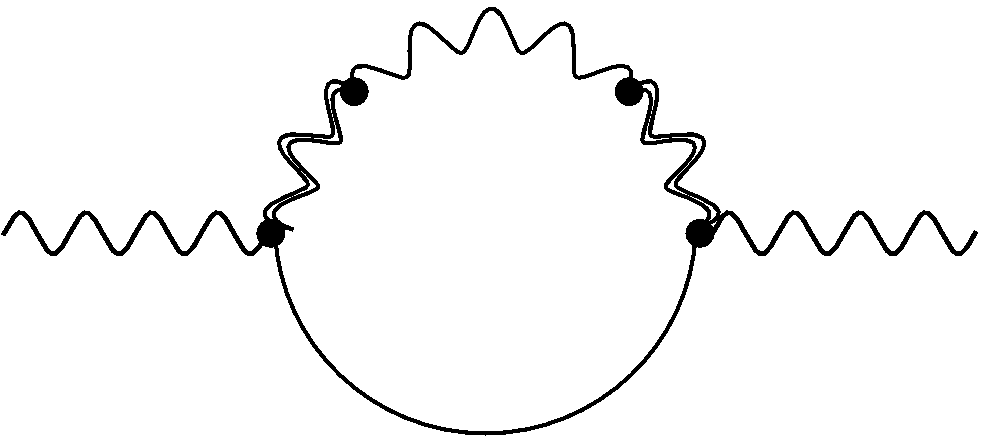}
\hspace{0.6cm}
\includegraphics[width=0.28\textwidth]{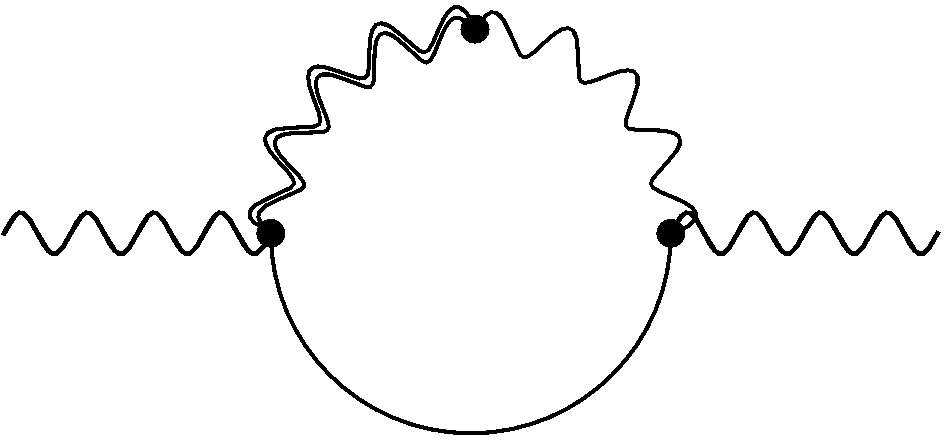}
\\[0.5cm]
\hspace{0.2cm}
\includegraphics[width=0.28\textwidth]{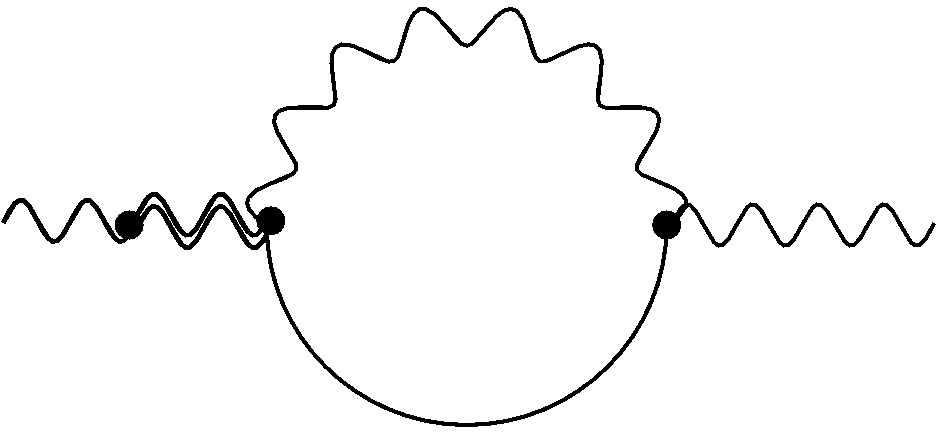}
\hspace{0.6cm}
\includegraphics[width=0.28\textwidth]{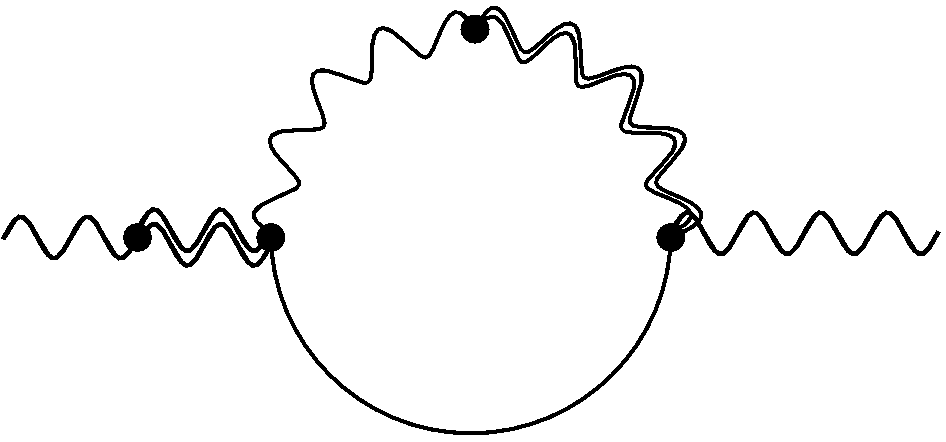}
\\[0.5cm]
\includegraphics[width=0.28\textwidth]{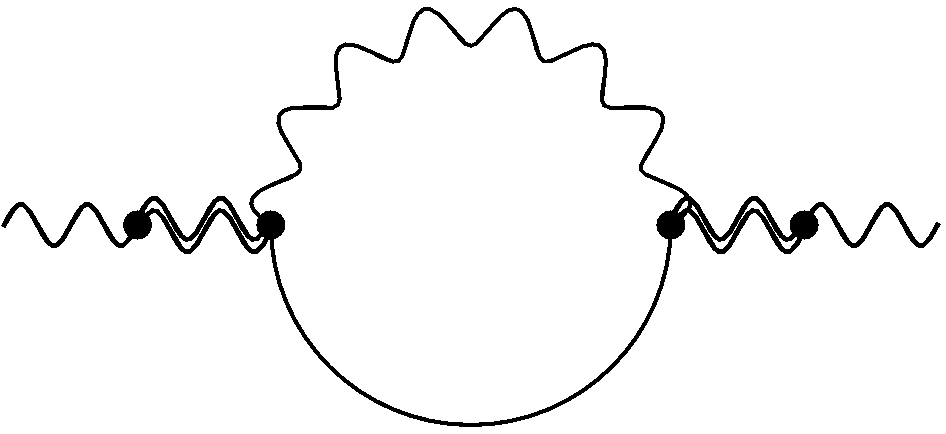}
\hspace{0.4cm}
\includegraphics[width=0.28\textwidth]{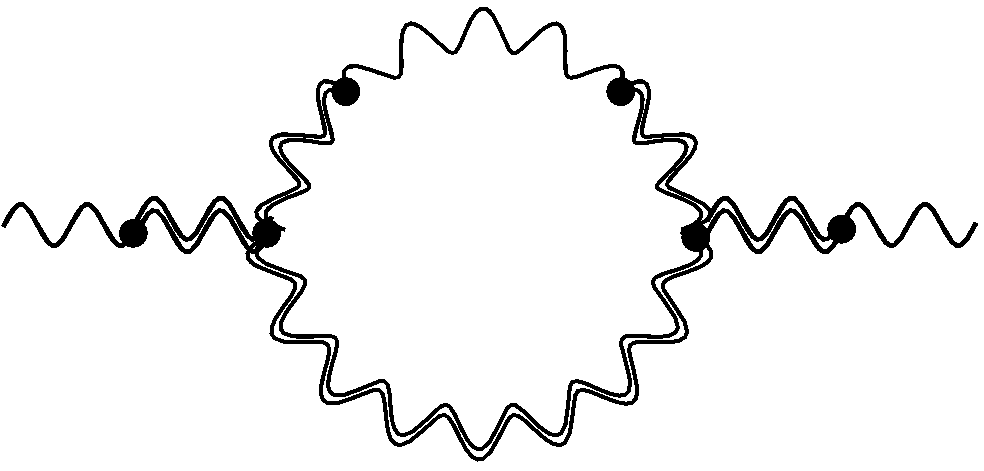}
\hspace{0.4cm}
\includegraphics[width=0.28\textwidth]{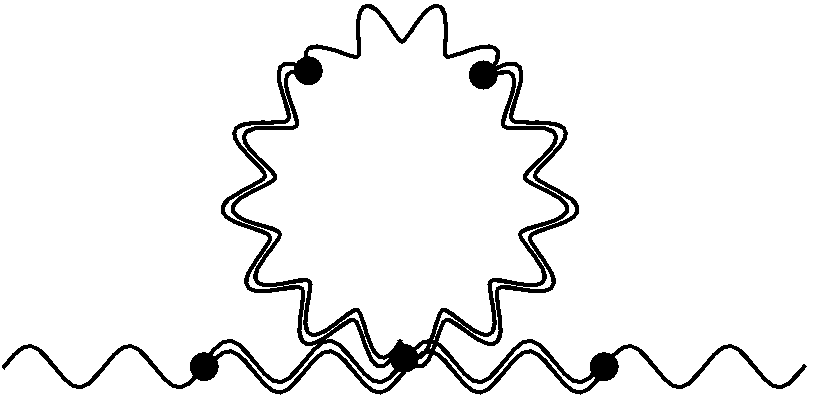}
\end{center}
\caption{
One-loop diagrams with $\rho$ exchange contributing to the $\langle W^1 W^1 \rangle - \langle W^3 W^3 \rangle$ Green function. 
External (internal) wavy lines denote the elementary $W$ ($B$) field, while continuous lines stand for the NG bosons ($\pi^{\hat a}$ and $\eta$).
The diagrams obtained by crossing the second, third and fourth one are not shown for simplicity.
}
\label{fig:altT_loops}
\end{figure}

As mentioned in the main text, the coefficient $\tilde c_T$ can be also extracted by matching the combination 
$\langle W^1 W^1 \rangle - \langle W^3 W^3 \rangle$ in the full and effective theories.
The relevant 1-loop diagrams are shown in Figs.~\ref{fig:altT_onlyNGbosons} and~\ref{fig:altT_loops} for the full theory ($\rho$ + NG bosons), and in 
Fig.~\ref{fig:altT_onlyNGbosons} for the low-energy theory of NG bosons.
%
%
Some of the diagrams have subdivergences associated with the renormalization of the $\rho$ propagator and of the $\rho-W$ mixing. The corresponding
counterterms in the unitary gauge are $(\Tr[\bar\rho^r_\mu \chi])^2$, $\Tr[d_\mu\chi] \Tr[\bar\rho^r_\mu \chi]$ and $\Tr[\bar\rho^L_\mu\chi]\Tr[\bar\rho^R_\mu\chi]$,
where $r = L,R$ and $\bar\rho^r_\mu \equiv \rho^r_\mu - E^r_\mu$.
The contribution of these counterterms, however, 
cancels out when summing all the diagrams. The overall divergence of the 
$\langle W^1 W^1 \rangle - \langle W^3 W^3 \rangle$ Green function is thus removed by 
the single counterterm $(\Tr[d_\mu \chi])^2$, as required to reproduce the calculation of $\tilde c_T$ through $\langle \pi^1 \pi^1 \rangle - \langle \pi^3 \pi^3 \rangle$. 
By matching the low-energy and full theories one obtains Eq.~(\ref{eq:matchingcT}).  A further check of the calculation follows from the fact that in the limit
$a_{\rho_L} = a_{\rho_R} = 1/\sqrt{2}$ the  counterterms combine into the $(SO(5)\times SO(5)_H)$-invariant operator of Eq.~(\ref{eq:OTsymm}).
In this limit the 1PI divergence vanishes, and the only divergent contribution to $\langle W^1 W^1 \rangle - \langle W^3 W^3 \rangle$ comes from subdiagrams.

\section{Results for a single $\rho$}
\label{sec:singlerho}

In a theory with a single spin-1 resonance, either $\rho_L$ or $\rho_R$, the RG evolution and matching conditions for $c_3^+$ and $c_T$ are respectively
(neglecting 1-loop contributions from $\alpha_{1,2}$)
\begin{align}
\mu \frac{d}{d\mu} c_3^+(\mu) & =   \frac{1}{192\pi^2} \left[  \frac{5}{4} + \frac{1}{4} a_{\rho}^2 (2 a_{\rho}^2 -7) \right]  \\[0.3cm]
\mu \frac{d}{d\mu} c_T(\mu) & = - \frac{3}{64\pi^2} \left(  1 - \frac{3}{4} a_{\rho}^2 \right)  
\end{align}
and
\begin{align}
\tilde c_3^+(\mu) & =  c_3^+(\mu)  -\frac{1}{2} \left( \frac{1}{4g_{\rho}^2} - \alpha_{2}  \right) 
    + \frac{1}{192\pi^2} \bigg[ \, \frac{3}{4} (a_{\rho}^2 + 28) \log\frac{\mu}{m_{\rho}}  + 1 + \frac{41}{16} a_{\rho}^2  \bigg] \\[0.3cm]
\tilde c_T(\mu) & =  c_T(\mu) - \frac{9}{256\pi^2} \bigg[ a_{\rho}^2  \log\frac{\mu}{m_{\rho}}  + \frac{3}{4} a_{\rho}^2 \bigg]\, .
\end{align}
The $\beta$-functions of $c_{2W}$ and $c_{2B}$ vanish. In a theory with only $\rho_L$ one has the matching conditions
\begin{align}
\begin{split}
\tilde c_{2W}(\mu) = & \, c_{2W}(\mu) - \frac{1}{2g_{\rho_L}^2 m_{\rho_L}^2} (1 - 2 \alpha_{2L} g_{\rho_L}^2)^2 \\
 & \, + \frac{1}{96\pi^2 m_{\rho_L}^2} \left[ 
  77 \log\frac{\mu}{m_{\rho_L}} + \frac{46}{5} - \frac{27}{32} a_{\rho_L}^2  \tan^2\frac{\theta}{2}  \right] 
\end{split} \\[0.3cm]
\tilde c_{2B}(\mu) = & \, c_{2B}(\mu)  \, , 
\end{align}
while only a $\rho_R$ gives
\begin{align}
\tilde c_{2W}(\mu) = & \, c_{2W}(\mu) \\[0.3cm]
\begin{split}
\tilde c_{2B}(\mu) = & \, c_{2B}(\mu) - \frac{1}{2g_{\rho_R}^2 m_{\rho_R}^2} (1 - 2 \alpha_{2R} g_{\rho_R}^2)^2 \\
 & \, + \frac{1}{96\pi^2 m_{\rho_R}^2} \left[ 
    77 \log\frac{\mu}{m_{\rho_R}} + \frac{46}{5} - \frac{27}{32} a_{\rho_R}^2  \tan^2\frac{\theta}{2}  \right] \, .
\end{split} 
\end{align}


\end{document}